%% file: multiaff_HottenrottLawsonRose.tex
\documentclass[12pt]{article}

\input{header}
\input{multiaff_statistics}

\date{January 2021}
\title{The Rise of Multiple Institutional Affiliations in Academia\thanks{
We are grateful to Stefano Baruffaldi and seminar participants at Max Planck Institute for Innovation and Competition, the Leibniz Center for Science and Society (LCSS) as well as the Ludwig Maximilian University Munich for valuable feedback. Carolin Formella, Iliana Radeva, Nurzhan Sapargali and Kaan Uctum provided valuable research assistance.}}

\author[1]{Hanna Hottenrott} 
\author[2]{Michael E. Rose}
\author[3]{Cornelia Lawson}
\affil[1]{{\small Technical University Munich, 80333 Munich, Germany, Tel: +498928928148, E-mail: hanna.hottenrott@tum.de (corresponding author)}}
\affil[2]{{\small Max Planck Institute for Innovation and Competition, Marstallplatz 1, 80532 Munich, Germany, Michael.Rose@ip.mpg.de}}
\affil[3]{{\small Alliance Manchester Business School, University of Manchester, Manchester M15 6PB, United Kingdom, cornelia.lawson@manchester.ac.uk}}

\begin{document}
\baselineskip20pt
\maketitle
\begin{singlespace}
\noindent
\textbf{Abstract} \\ 
This study provides the first systematic, international, large-scale evidence on the extent and nature of multiple institutional affiliations on journal publications. Studying more than 15 million authors and 22 million articles from 40 countries we document that: In 2019, almost one in three articles was (co-)authored by authors with multiple affiliations and the share of authors with multiple affiliations increased from around 10\% to 16\% since 1996. The growth of multiple affiliations is prevalent in all fields and it is stronger in high impact journals. About 60\% of multiple affiliations are between institutions from within the academic sector. International co-affiliations, which account for about a quarter of multiple affiliations, most often involve institutions from the United States, China, Germany and the United Kingdom, suggesting a core-periphery network. Network analysis also reveals a number communities of countries that are more likely to share affiliations. We discuss potential causes and show that the timing of the rise in multiple affiliations can be linked to the introduction of more competitive funding structures such as “excellence initiatives” in a number of countries. We discuss implications for science and science policy.  \\
\end{singlespace}

\newpage
\section{Introduction}
Institutions have an important role in academic research. They impact researchers' work as they control access to resources, networks and research infrastructure, and thus partially determine scientific discovery \citep{Stephan2012HowScience}. Institutional affiliation moreover affects the value ascribed to individual researchers through institutional prestige, with consequences for research and career trajectories. Science policy has further lifted the value assigned to institutions, through the use of domestic and international rankings and the introduction of performance-based institutional funding mechanisms which it believes will encourage greater research performance \citep{Salmi2916}. 
We suggest that as a consequence of the inflated importance of affiliations, more academics are now affiliated to multiple institutions and are reporting these on their academic work. Yet, so far multiple affiliations (or co-affiliations), where researchers are formally attached to more than one institution at the same time \citep{KATZ19971,Hottenrott2017AUK}, are largely unexplored. A three country and three field study conducted by \cite{Hottenrott2017AUK} provides some first evidence on the extent and structure of multiple affiliations, showing an increase in the three countries and scientific fields under study. \cite{matveeva2019} reported an increase also for Russia. In this article we expand on this prior research. In an analysis of a set of 40 countries over a 24 year time period we focus on international differences in multiple affiliations, which are important to understand potential drivers and consequences of this phenomenon.

There are a number of reasons why multiple affiliations may occur and which may explain why they differ internationally. For one, they may be driven by individual research trajectories. A prestigious affiliation can serve as a "mechanism for cumulative advantage" \citep{Way2019} and researchers may seek out affiliations to institutions outside their main employment to gain or maintain access to resources and networks. For instance, many science systems have prominent public research organisations (PROs) that contribute substantially to research production, but not teaching. The prestige and high level of research infrastructure available at these PROs make them attractive for academics at universities. Moreover, in a number of countries PRO affiliation is encouraged or linked to professorial posts. For instance, the Chinese Academy of Science which offers affiliation to leading Chinese scientists and has more than 50,000 members is listed as the top publishing institution on Nature Index\footnote{See \href{https://www.natureindex.com/}{www.natureindex.com}.}, a database of author affiliations on articles in selected top journals \citep{Li2016}.
Further, in internationalised research with high levels of mobility, diaspora networks form \citep{meyer2006diaspora,MIGUELEZ2020}. Such networks may be informal or formal, and can include association with learned societies, visiting positions, or dedicated diaspora initiatives in the home country \citep{Baruffaldi2012}. Institutions and mobile researchers may formalise these linkages in multiple affiliations to enable knowledge exchange and curb the effects of brain drain. Such international linkages could be particularly important for the Global South to redress inequalities in research production \citep{LANGA2018}.

The importance of domestic research and diaspora differ between countries and therefore the prevalence and types of multiple affiliations differ too. \cite{Hottenrott2017AUK}, for instance, find in a sample of articles drawn from the Web of Science that international co-affiliations are dominant for academic authors in the United Kingdom, while cross-sector co-affiliations are more common in Japan. This reflects the importance of internationalised communities and domestic PROs respectively, and providing first clues towards the international differences we may observe. 

Any increases in multiple affiliations, however, will not be entirely due to individual research trajectories. Rankings and research assessments, mentioned above, can also be a cause for the increase of multiple affiliations, and could provide an explanation for why they may differ internationally. Specifically, any country differences in multiple affiliations may be due to the shift in some countries towards initiatives designed to improve their competitiveness in international research. Some countries recently implemented major reforms in the resource allocation processes within the research sector, moving towards performance-based allocation of funding through schemes such as research excellence initiatives (ExIns). These initiatives aim to accelerate the transformation of higher education and to boost the research capacity and productivity of academic institutions \citep{Salmi2916}. This is achieved via block grants being made available to selected institutions or for the establishment of new centres of excellence such as in the case of Germany, or via research evaluation such as the Research Excellence Framework in the United Kingdom \citep{Salmi2916,GEUNA2016260}. ExIns thus introduce competition and performance-based funding elements into higher education systems, which are said to increase performance \citep{Aghion2014}, and are arguably the most important factor for universities to move up in global rankings \citep{benito2019funding}. Sources of such performance change other than the monetary investments are the concentration of resources, focus on measurable research outputs, and orientation towards international research agendas and publication outputs, which contribute towards higher international visibility \citep{Salmi2916}. There is substantial existing evidence on the impact of ExIns on publication output, for example from China’s 985 project \citep{ZHANG2013765,Zong2019}, Russia's 5–100 project \citep{Agasisti2020,turko2016influence}, Germany's Excellence Initiative \citep{menter2018search,CIVERA2020104083}  or Taiwan's World Class University project \citep{Fu2019}, which provide evidence of positive, but also mixed performance effects of these initiatives.

Multiple affiliations may increase as a consequence of ExIns for a number of reasons. ExIns are closely linked to research assessment and an expectation of performance improvement. Universities in an attempt to upgrade quickly, may buy in external talent instead of building up capacity locally or restructure their activities. In Russia, for instance, a performance effect from ExIns was quickly observed, as was an increase in multiple affiliations \citep{matveeva2019}, suggesting that universities hired external talent. Such co-affiliations were potentially aimed at a quick increase in rankings based on bibliometric data. In France, in turn, research and teaching activities needed to be concentrated "under one roof and one name" to build greater visibility in international rankings \citep{Paradeise2018}, achieved via a closer integration of universities and CNRS research centres. In this context multiple affiliations are an outcome of formalised or institutionalised forms of collaborations \citep{hicks1996spp}, which have also emerged as an explicit goal of ExIns in some countries, such as Germany's Clusters of Excellence \citep{bornmann2016promotion,Froumin2015}.

There is also some evidence that ExIns aimed at elite institutions may not only improve the international ranking of 'excellence' universities \citep{Salmi2916,benito2019funding}, but that they can lift other, unfunded institutions \citep{Fu2019,Agasisti2020,CIVERA2020104083}. The increase in stratification associated with ExIns and the concentration of resources make emerging elite institutions very attractive for outside researchers. Academics at lower ranked institutions may thus seek access to the resources available there. \cite{Hamann2018} for instance shows more inward mobility into high-rank institutions following a research evaluation round in the United Kingdom. In addition, the metrics that determine resource allocation also serve as a reference point for lower ranked or unfunded institutions, who may adopt similar strategies in an attempt to improve their position in future evaluation or funding rounds. 

Despite first tentative findings, there is need for a better understanding of the extent and nature of multiple affiliations and how they may relate to research excellence initiatives in different countries. In particular, there is a lack of a complete comparison of multiple affiliation trends within research nations, especially those that implemented ExIns. To fill this gap, this study sets out to answer the following three research questions:

\begin{itemize}
\item RQ1: How have multiple affiliations evolved over time in different countries and scientific fields?
\item RQ2: Do we observe any patterns in the geographic location and in the sectors of co-affiliations? 
\item RQ3: Do country differences in the evolution of multiple affiliations relate to policy changes in funding allocation (ExIns) in different countries?
\end{itemize}

 To investigate the change of multiple affiliations over time we present a large-scale bibliometric analysis that makes use of the affiliation information of \NOfAuthorsUnique\ different authors located in 40 countries. All data originate from Scopus. We use \NOfArticlesUnique\ research articles published between 1996 and 2019 that are representative for 26 distinct scientific fields. 
 
 The use of publication affiliation data to investigate multiple affiliations is not only appropriate but also highly relevant, as it is used frequently to assign research achievements to scientific institutions, such as with university rankings \citep{GEUNA2016260}. For example, the Times Higher Education ranking since 2016 uses Scopus data for its popular THE World University Rankings.\footnote{See \href{https://blog.scopus.com/posts/times-higher-education-choose-scopus-data-for-its-world-university-ranking}{blog.scopus.com}.} This is not without problem. If not accounting for multiple affiliations, each document counts as often as there are distinct affiliations reported on the publication. A simple example shows that distortions arise easily. Consider university $U$ with 4 researchers, research institute $R$ with 2 researchers, and College $C$ with 3 researchers. If each researcher publishes one article, the ranking obtains as $U > C > R$. If however two of $U$'s researchers have a multiple affiliation with $R$, the ranking changes to $U > R > C$. As such the reliance of research rankings on bibliometric data could create incentives for institutions to offer affiliations to researchers primarily employed elsewhere. Evidence comes from \cite{Bornmann2015}, who analyse the affiliations of highly cited authors and find that Saudi Arabia emerges high in the country ranking when secondary affiliations are counted. As a consequence, research discoveries may be assigned to places where they did not necessarily originate \citep{Xin2006,Bhattacharjee2011}. In the example above, did $R$ or did $U$ enable the research of the two researchers that hold multiple affiliations at both $R$ and $U$, or did both contribute? While this is not a question we are able to answer, it is one that emphasises the relevance of investigating multiple affiliations.

Our study thus contributes towards a better understanding of multiple affiliations which will also inform how we think about creative places and will thus ultimately impact science policy \citep{deRijcke2015,hicks1996spp}.

\section{Methods}
\subsection{Data collection}
We base our analysis of affiliations on documents published between 1996 and 2019.\footnote{We use only research articles, reviews, notes, conference proceedings, in-press articles or short articles, as classified by Scopus.} Counting affiliations from published articles has the advantage of being available at a large scale and with high coverage (as opposed to, e.g. CVs or university websites). We derive all data from Scopus which has three important advantages over other indexing systems: comprehensive coverage of scientific articles \citep{MongeonPaul-Hus2016}, disambiguation of authors and their affiliations along with the assignment of unique Author IDs, and availability of additional information on institution addresses (e.g. country and organisation type). 

We study all 26 scientific fields as identified by the All Science Journal Classification (ASJC) codes excluding the category 'Multidisciplinary'. Table \ref{tab:usable_articles} lists these together with the number of papers and authors used. The analysis focuses on 40 countries: All OECD countries as of 2019 excluding Latvia, Luxembourg and Iceland (because they host too few universities), and a group of non-OECD science producing countries consisting of Argentina, China, Romania, Russia, Singapore, South Korea, South Africa and Taiwan.

We obtain the set of articles by field through the set of journals representing that field. To obtain these journals, we make use of the Scimago journal ranking\footnote{See \url{https://www.scimagojr.com/}.}, which is also based on the Scopus database. For each field we first remove journals with fewer than three citations in the previous three years in any year $t$ and we remove sources with at least five years of coverage in the 1996-2019 period. Using the Scimago Journal Impact Factor (SJR) we drop the lower 50\% of the journal quality distribution. The upper 50\% thus represent the set of journals that we study. If journals are assigned to multiple fields, we also assign articles published therein to multiple fields. Forty-three percent of journals list more than one field and on average each journal belongs to 1.63 fields, with eight fields being the maximum. We then retrieve bibliometric information for all articles published in these journals during the 1996-2019 period using the `pybliometrics' module developed by \citet{Rose2019}.\footnote{The data was downloaded between March 2020 and September 2020.} This stage excludes articles authored by either anonymous authors or a collaborative unit.

Next, we remove observations with missing affiliations\footnote{Affiliations are missing in the Scopus database when, for example, affiliation information is too hard to parse automatically.} and mark an article `usable' if it provides affiliation information for at least one author. The share of articles with 'usable' affiliations is close to 100 percent in the majority of fields and increasing over time (Figure \ref{fig:useable_articles}). Articles are further removed if: they are not research-type articles (e.g. editorials, reviews, etc.), the author and/or affiliation information is completely absent, or none of the article authors is from the selected set of countries. Unique authors are then identified based on their Scopus Author ID.\footnote{When generating author profiles, the Scopus algorithm is conservative and prefers "split profiles" over "merge profiles" \citep{Moed2013StudyingScopus}, i.e. it rather casts too many profiles for the same researcher than lumping publications of many different researchers into one profile. Accordingly, \citet{Baas2020ScopusStudies} estimate the precision of Scopus author profiles (the absence of documents that belong to someone else) equal to 98.1\% and recall (the absence of documents that belong to this author) equal to 94.4\%. Precision and recall do however correlate with origin of authors. This is likely the reason behind the high share of Chinese authors (read: author profiles) in our samples (Table \ref{tab:articlesauthors_country}). To check for any bias introduced by this, we compare figures for the full sample to a subsample excluding China and see only very minor differences. To illustrate this, Figure \ref{fig:multiaff_fields-countryfield_nochina} plots the share of authors with multiple affiliations by field, excluding China. The difference to Figure \ref{fig:multiaff_fields-countryfield} that shows the share of authors with multiple affiliations by field for the full sample is minimal.}

Table \ref{tab:usable_articles} outlines this sampling and selection strategy. For each field we list the total number of journals available, the number of journals sampled following the selection described above, the number of articles sampled and finally used, and the number of authors on these articles. The largest number of journals is selected in the field of Medicine (3,205) and the fewest in Dentistry (86). Articles can appear several times if the journal in which they are published is categorized into two or more different research fields. Authors also appear several times if they publish in multiple fields. In other words, we usually study author-field pairs to avoid assigning an author to a single field based on our own judgement.

\begin{table}[ht]
 \caption{Number of journals, authors and articles used in the study, by field \label{tab:usable_articles}}
 \centering
 {\scriptsize
\begin{tabular}{lrrr|rrrrr|r}
\toprule
{} & \multicolumn{3}{c}{Journals} & \multicolumn{5}{c}{Articles} &     Authors \\
{} &    Total & Coverage & Sampled &      Sampled & research-type &      Usable &        Used & Share &        Used \\
Field &  & > 5 years &  &       &  &  &  & (in \%) & \\
\midrule
Arts/Humanities   &    3,451 &              3,027 &   1,503 &  1,057,138 &       998,161 &   890,076 &    825,418 &        78.08 &    849,296 \\
Biochemistry      &    1,973 &              1,789 &     894 &  3,840,559 &     3,657,624 & 3,616,530 &  3,459,702 &        90.08 &  4,460,688 \\
Biology           &    2,018 &              1,833 &     915 &  2,395,094 &     2,338,967 & 2,322,030 &  2,101,027 &        87.72 &  2,537,858 \\
Business          &    1,197 &              1,099 &     548 &    504,956 &       484,784 &   471,073 &    432,109 &        85.57 &    380,911 \\
Chem. Engineering &      551 &                490 &     245 &  1,313,061 &     1,286,950 & 1,280,504 &  1,141,440 &        86.93 &  1,655,801 \\
Chemistry         &      787 &                747 &     373 &  2,864,867 &     2,813,743 & 2,802,557 &  2,546,212 &        88.88 &  2,738,343 \\
Computer Sci.     &    1,441 &              1,300 &     650 &  1,167,238 &     1,131,391 & 1,123,201 &  1,026,233 &        87.92 &  1,204,346 \\
Decision Sci.     &      339 &                306 &     152 &    238,521 &       232,554 &   230,957 &    215,060 &        90.16 &    205,877 \\
Dentistry         &      193 &                172 &      86 &    178,588 &       165,899 &   159,827 &    133,391 &        74.69 &    169,910 \\
Economics         &      915 &                832 &     416 &    389,043 &       379,147 &   371,509 &    349,820 &        89.92 &    239,495 \\
Energy            &      388 &                338 &     169 &    722,958 &       713,002 &   710,319 &    626,275 &        86.63 &    919,263 \\
Engineering       &    2,436 &              2,216 &   1,106 &  3,457,267 &     3,389,889 & 3,360,381 &  3,005,072 &        86.92 &  3,270,398 \\
Environ. Sci.     &    2,437 &              2,218 &   1,108 &  3,470,738 &     3,403,196 & 3,373,675 &  3,017,574 &        86.94 &  3,274,732 \\
Health            &      501 &                459 &     229 &    474,952 &       441,382 &   429,627 &    406,509 &        85.59 &    657,647 \\
Immunology        &      539 &                490 &     245 &    956,989 &       906,776 &   895,494 &    842,936 &        88.08 &  1,453,548 \\
Materials Sci.    &    1,091 &                985 &     491 &  2,761,531 &     2,719,737 & 2,708,951 &  2,433,903 &        88.14 &  2,333,041 \\
Mathematics       &    1,338 &              1,249 &     622 &  1,231,745 &     1,210,597 & 1,201,599 &  1,098,621 &        89.19 &    829,208 \\
Medicine          &    7,086 &              6,414 &   3,205 & 10,166,641 &     9,158,301 & 8,908,234 &  8,283,039 &        81.47 &  8,104,289 \\
Neuroscience      &      552 &                494 &     247 &    899,634 &       842,383 &   831,151 &    803,158 &        89.28 &  1,183,954 \\
Nursing           &      598 &                551 &     275 &    508,495 &       462,258 &   444,547 &    415,122 &        81.64 &    693,566 \\
Pharmaceutics     &      727 &                668 &     334 &  1,057,268 &     1,015,266 & 1,003,096 &    898,004 &        84.94 &  1,636,838 \\
Physics           &    1,010 &                945 &     472 &  3,229,665 &     3,174,338 & 3,163,537 &  2,893,511 &        89.59 &  2,437,930 \\
Planetary Sci.    &    1,073 &                988 &     494 &  1,281,412 &     1,254,875 & 1,249,080 &  1,178,801 &        91.99 &    951,187 \\
Psychology        &    1,119 &              1,042 &     519 &    704,218 &       675,294 &   664,295 &    645,091 &         91.6 &    653,761 \\
Social Sci.       &    5,552 &              4,899 &   2,444 &  1,982,561 &     1,885,055 & 1,783,644 &  1,656,468 &        83.55 &  1,488,451 \\
Veterinary        &      226 &                208 &     104 &    287,436 &       269,724 &   255,368 &    220,001 &        76.54 &    334,073 \\
Total (Unique)            &   23,367 &             21,106 &  10,976 &            &               &           & 22,198,910 &              & 15,234,353 \\
\bottomrule
\end{tabular}
}
 \justify \small{\textit{Notes:} The table reports steps of the sampling and selection process. For each field, we select all journals with at least 5 years of publication history in the 1996--2019 period according to Scopus. Of these we restrict to journals in the upper half of the SJR quality distribution as of 2019. For each journal, we deduct non-research type documents, documents with missing author and affiliation information (`Usable'), and finally documents where all authors are from countries outside our sampling frame. Column `Share' reports the share of `Used' documents over `Sampled' documents. Column `Authors Used' reports the number of unique authors of the `Used' documents.}
\end{table}

\subsection{Identifying multiple affiliations}
To identify multiple affiliations we use authors' affiliation information on articles in the Scopus database.\footnote{Appendix \ref{sec::data} details two refinements to the affiliation information aimed at avoiding to falsely classifying an additional address as affiliation information.} An article is considered to be a multiple affiliation article if at least one of the authors lists multiple affiliations. When we study author-year or author-year-field observations, an author is considered to have multiple affiliations if they list more than one affiliation on at least one of their articles in that year and field. We thus consider two different measures of multiple affiliation, by article and by author.

We assume the first listed affiliation to be the main affiliation, and hence take the country of the first affiliation as "home" country of the author. Table \ref{tab:articlesauthors_country} shows the number of authors and articles by country of first affiliation for the full dataset. Authors are marked as having an international affiliation if we observe a second affiliation in a country that is different from the "home" country. 

\FloatBarrier
\subsection{Computing shares of authors with multiple affiliations\label{aggregation}}
When computing shares of authors with multiple affiliations by country and by field, there are in general two ways to do so. One way accounts for the different distribution of articles across fields in different countries (or the distribution of articles across countries), the other does not. Formally, let $a$ denote the number of authors and $a^{M}$ denote the number authors with multiple affiliations in country $c$ in field $f$ and year of publication $t$. Throughout the analysis we use the average country-year share of authors with multiple affiliations defined as:

\begin{equation}\label{eq:share-country}
   s_{c,t} = \frac{\sum a^{M}_{c,t}}{\sum a_{c,t}}.
\end{equation}

Likewise, when aggregating over fields, we can use the average share of authors with multiple affiliations per field as: 

\begin{equation}\label{eq:share-field}
   s_{f,t} = \frac{\sum a^{M}_{f,t}}{\sum a_{f,t}}.
\end{equation}

Alternatively to averaging directly across countries or fields, we can average in two steps by first calculating the country-field level average share: 

\begin{equation*}\label{eq:share}
   m_{c,f,t} = \frac{\sum a^{M}_{c,f,t}}{\sum a_{c,f,t}}.
\end{equation*}

and then take the average over these values either by country or field: 

\begin{equation}\label{eq:share-countryfield}
    \overline{s}_{c,t} = \frac{m_{c,f,t}}{\sum_c m_{c,f,t}}.
\end{equation}

\begin{equation}\label{eq:share-fieldcountry}
   \overline{s}_{f,t} = \frac{m_{c,f,t}}{\sum_f m_{c,f,t}}.
\end{equation}

Qualitatively both methods paint very similar pictures as shown in a comparison between Figure \ref{fig:multiaff_global}(A) and Figure \ref{fig:multiaff_fields-countryfield}.

\FloatBarrier

\section{Results}
\subsection{The evolution of multiple affiliations}
\begin{figure}[ht]
    \centering
    \caption{Authors with multiple affiliations
    \label{fig:multiaff_global}}
    \includegraphics[width=0.8\textwidth]{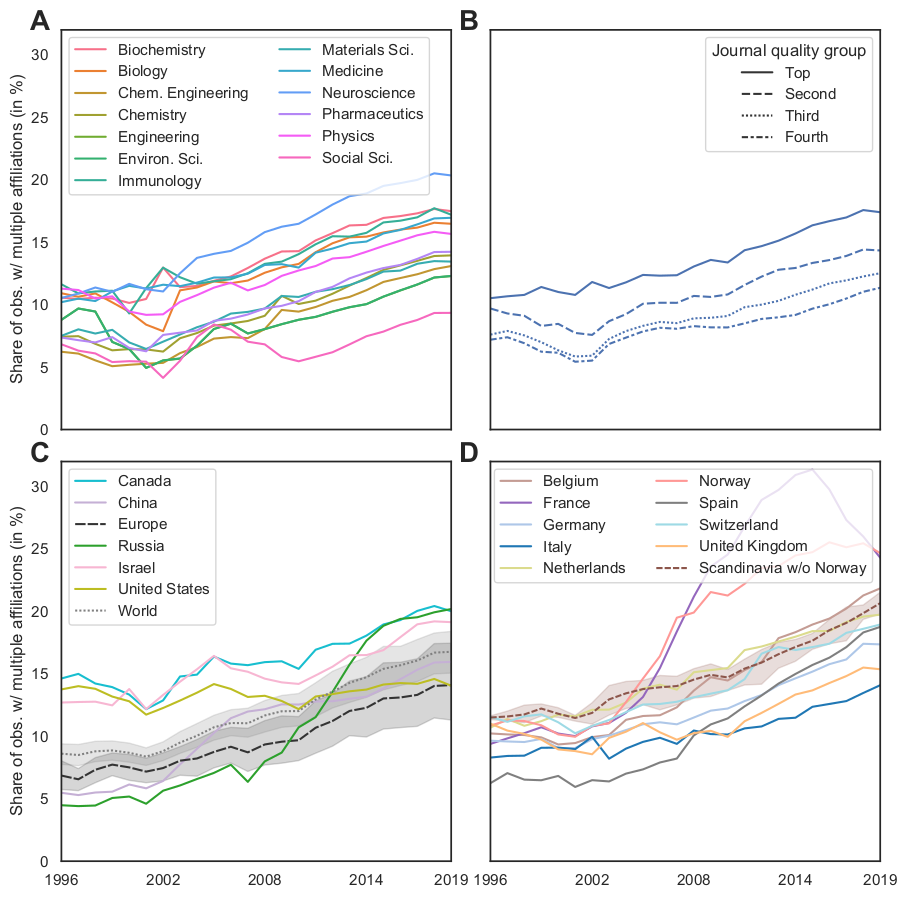}
    \justify \small{\textit{Notes:} (A) Share of authors with multiple affiliations by field. Authors are assigned to fields based on the fields to which their article’s journal is assigned. Authors can appear multiple times if they publish in different years or in different fields. See Table \ref{tab:multiaff_authors_share-field} for precise values. (B) Share of authors with multiple affiliations by journal quality group. A journal quality group corresponds to a quartile in the journal impact factor distribution of the journals included in our study (in cases where a journal is ranked in multiple fields, the higher quartile applies). See Table \ref{tab:multiaff_authors_share-quality} for precise values. (C/D) Share of authors with multiple affiliations by country. Authors are assigned to the country of their first listed affiliation per publication. See Table \ref{tab:multiaff_authors_share-country} for precise values.}
\end{figure}

We firstly investigate the evolution of multiple affiliations over time to investigate RQ1. The results presented in Figure \ref{fig:multiaff_global} and Table \ref{tab:multiaff_articles_share-field} show that multiple affiliations are a global phenomenon and that they are on the rise. Figure \ref{fig:multiaff_global} shows that they have become more prevalent in all scientific fields, all journal quality groups, and the vast majority of countries. From Table \ref{tab:multiaff_articles_share-field}, which reports the share of articles with at least one author with multiple affiliations by year and field, we see that, while on average about one quarter of articles qualify as having multiple affiliations, this figure was lowest in 2001 with a share of 14.7\% and highest in 2018 with 32.5\%. While increasing in all fields, in 2019 the share of articles with at least one author with multiple affiliations is highest in Neuroscience (49.3\%) and lowest in Arts/Humanities (13.6\%). 

\begin{table}
 \caption{Share of articles with multiple affiliations by field over time (in \%)
 \label{tab:multiaff_articles_share-field}}
 \resizebox{\columnwidth}{!}{%
 \centering
\begin{tabular}{lrrrrrrrrrrrrrrrrrrrrrrrr|r}
\toprule
{} &  1996 &  1997 &  1998 &  1999 &  2000 &  2001 &  2002 &  2003 &  2004 &  2005 &  2006 &  2007 &  2008 &  2009 &  2010 &  2011 &  2012 &  2013 &  2014 &  2015 &  2016 &  2017 &  2018 &  2019 &  Average \\
\midrule
Arts/Humanities   &  10.2 &   9.8 &  10.0 &   9.3 &   8.5 &   6.9 &   7.3 &  10.5 &  11.3 &  12.1 &  12.0 &  11.2 &  11.5 &  10.4 &   9.3 &   9.4 &  10.0 &  10.1 &  10.9 &  11.3 &  11.9 &  12.6 &  13.6 &  13.6 &              10.6 \\
Biochemistry      &  22.8 &  23.1 &  23.5 &  23.9 &  21.4 &  20.4 &  22.0 &  26.3 &  28.3 &  29.1 &  29.4 &  30.8 &  32.0 &  32.4 &  32.5 &  34.6 &  36.6 &  38.2 &  39.2 &  40.6 &  41.7 &  42.4 &  43.5 &  43.4 &              31.6 \\
Biology           &  19.8 &  19.5 &  20.1 &  19.3 &  17.1 &  15.2 &  15.3 &  23.1 &  23.8 &  25.2 &  24.9 &  25.1 &  26.4 &  27.0 &  28.3 &  31.2 &  33.1 &  34.4 &  35.1 &  36.1 &  36.8 &  37.1 &  38.1 &  37.9 &              27.1 \\
Business          &   7.3 &   8.3 &   7.7 &   6.3 &   6.3 &   6.5 &   4.9 &   6.1 &   8.7 &  11.0 &  11.5 &   9.7 &   9.4 &   7.9 &   6.8 &   7.9 &   9.1 &  10.3 &  11.1 &  11.9 &  12.5 &  13.3 &  14.5 &  15.1 &               9.3 \\
Chem. Engineering &  13.0 &  12.6 &  11.7 &  10.5 &  11.0 &  11.1 &  11.1 &  13.8 &  14.9 &  16.6 &  16.7 &  16.8 &  18.3 &  19.7 &  21.1 &  22.8 &  24.5 &  25.4 &  27.0 &  28.6 &  29.6 &  30.8 &  32.1 &  32.5 &              19.7 \\
Chemistry         &  14.6 &  14.6 &  13.6 &  12.2 &  12.5 &  12.4 &  12.6 &  15.7 &  16.9 &  18.2 &  18.4 &  19.2 &  20.1 &  21.6 &  22.4 &  23.9 &  25.4 &  27.0 &  28.3 &  30.0 &  31.0 &  32.2 &  33.5 &  33.7 &              21.3 \\
Computer Sci.     &  12.3 &  13.3 &  13.0 &  10.0 &   9.8 &   9.3 &   9.6 &  10.0 &  11.1 &  14.5 &  15.0 &  13.0 &  13.6 &  14.0 &  14.6 &  15.9 &  16.8 &  18.0 &  19.4 &  20.9 &  22.0 &  22.7 &  24.3 &  24.6 &              15.3 \\
Decision Sci.     &   8.1 &   9.3 &   9.8 &   6.6 &   7.4 &   8.6 &   6.2 &   7.9 &  11.2 &  13.1 &  12.2 &  11.5 &  11.5 &  11.8 &  10.3 &  12.1 &  13.3 &  14.5 &  15.7 &  16.0 &  17.5 &  18.7 &  20.1 &  20.4 &              12.3 \\
Dentistry         &  20.2 &  18.2 &  19.8 &  17.2 &  18.2 &  17.6 &  17.1 &  20.3 &  21.0 &  21.4 &  19.2 &  18.0 &  19.4 &  17.9 &  18.6 &  19.9 &  20.1 &  21.8 &  21.5 &  23.9 &  24.2 &  25.9 &  27.9 &  28.5 &              20.7 \\
Economics         &   8.7 &   7.4 &   7.7 &   8.0 &   8.4 &   8.6 &   8.0 &   9.0 &  12.2 &  13.1 &  12.8 &  13.0 &  12.8 &  12.1 &  11.9 &  13.2 &  14.7 &  15.1 &  16.3 &  16.9 &  16.6 &  16.6 &  18.0 &  17.6 &              12.4 \\
Energy            &  17.2 &  18.5 &  17.3 &   9.8 &  11.1 &  12.1 &   9.6 &  11.0 &  12.8 &  15.3 &  17.2 &  18.3 &  17.4 &  16.5 &  17.5 &  18.5 &  20.1 &  21.4 &  22.5 &  23.9 &  25.2 &  26.3 &  27.9 &  28.0 &              18.1 \\
Engineering       &  13.0 &  14.6 &  14.1 &  11.1 &  10.7 &   9.2 &  10.1 &  11.4 &  13.0 &  15.5 &  16.1 &  15.2 &  15.7 &  16.1 &  17.1 &  18.0 &  19.0 &  19.9 &  20.9 &  21.9 &  23.1 &  23.9 &  25.5 &  25.7 &              16.7 \\
Environ. Sci.     &  13.0 &  14.5 &  14.1 &  11.0 &  10.7 &   9.2 &  10.1 &  11.4 &  13.0 &  15.5 &  16.1 &  15.2 &  15.7 &  16.1 &  17.1 &  18.0 &  19.0 &  20.0 &  20.9 &  21.9 &  23.1 &  24.0 &  25.6 &  25.7 &              16.7 \\
Health            &  20.7 &  20.3 &  19.9 &  20.3 &  16.3 &  16.6 &  16.9 &  20.9 &  24.1 &  25.7 &  24.2 &  24.6 &  26.1 &  24.9 &  22.1 &  24.3 &  26.1 &  27.1 &  28.6 &  29.6 &  31.0 &  32.3 &  35.4 &  35.3 &              24.7 \\
Immunology        &  28.0 &  25.6 &  27.0 &  26.5 &  20.4 &  17.4 &  18.1 &  29.1 &  30.4 &  30.6 &  30.0 &  31.6 &  33.3 &  32.6 &  33.5 &  35.1 &  36.5 &  37.2 &  38.6 &  40.8 &  41.8 &  42.9 &  45.0 &  44.5 &              32.4 \\
Materials Sci.    &  13.9 &  14.4 &  14.2 &  14.2 &  12.4 &  12.6 &  14.0 &  16.0 &  17.5 &  18.6 &  19.3 &  19.9 &  20.5 &  21.4 &  22.9 &  24.2 &  25.3 &  26.3 &  27.6 &  29.1 &  29.4 &  30.7 &  31.7 &  31.8 &              21.2 \\
Mathematics       &  10.5 &  11.0 &  11.0 &  10.1 &   9.8 &   9.6 &   8.8 &  10.7 &  11.8 &  13.3 &  13.3 &  13.0 &  13.6 &  14.4 &  15.0 &  15.4 &  16.0 &  17.0 &  18.3 &  19.1 &  20.0 &  20.3 &  21.4 &  21.2 &              14.4 \\
Medicine          &  20.8 &  21.0 &  21.4 &  21.5 &  20.0 &  18.6 &  18.5 &  24.1 &  26.7 &  27.4 &  26.6 &  27.3 &  28.5 &  27.7 &  26.4 &  28.8 &  30.1 &  31.6 &  32.7 &  34.4 &  35.8 &  37.0 &  38.5 &  39.0 &              27.7 \\
Neuroscience      &  22.3 &  22.4 &  24.7 &  24.3 &  24.2 &  24.6 &  24.1 &  28.8 &  32.8 &  33.7 &  33.6 &  34.5 &  36.0 &  36.3 &  35.8 &  37.8 &  40.1 &  41.5 &  43.0 &  44.9 &  45.8 &  46.9 &  48.9 &  49.3 &              34.9 \\
Nursing           &  16.6 &  15.3 &  15.7 &  13.7 &  13.3 &  12.5 &  15.2 &  18.1 &  20.1 &  21.7 &  22.4 &  23.8 &  23.8 &  22.5 &  20.3 &  22.9 &  24.3 &  26.1 &  26.6 &  28.4 &  30.0 &  31.9 &  33.6 &  35.2 &              22.3 \\
Pharmaceutics     &  17.6 &  16.6 &  16.6 &  16.6 &  14.5 &  13.9 &  14.8 &  20.1 &  21.1 &  22.8 &  22.7 &  23.4 &  24.3 &  24.9 &  25.2 &  27.7 &  29.0 &  30.7 &  32.4 &  33.4 &  34.7 &  35.6 &  37.0 &  37.2 &              24.7 \\
Physics           &  18.9 &  19.4 &  18.3 &  19.3 &  17.3 &  17.4 &  16.7 &  20.6 &  22.0 &  22.9 &  23.2 &  22.8 &  23.8 &  25.1 &  26.6 &  27.9 &  29.3 &  30.2 &  31.3 &  32.8 &  34.0 &  34.8 &  35.8 &  35.6 &              25.3 \\
Planetary Sci.    &  22.4 &  23.5 &  22.2 &  20.9 &  18.2 &  16.9 &  16.6 &  18.3 &  21.6 &  26.5 &  26.0 &  26.4 &  26.7 &  27.8 &  29.2 &  31.2 &  32.6 &  33.4 &  35.2 &  37.1 &  38.7 &  40.0 &  41.9 &  42.4 &              28.2 \\
Psychology        &  17.5 &  15.4 &  15.5 &  12.8 &  13.0 &  13.3 &  14.4 &  18.1 &  22.6 &  23.9 &  23.4 &  20.8 &  19.6 &  17.7 &  17.0 &  19.4 &  20.8 &  23.4 &  24.7 &  26.6 &  26.8 &  27.6 &  29.0 &  28.6 &              20.5 \\
Social Sci.       &   8.4 &   8.1 &   7.9 &   7.1 &   6.9 &   6.5 &   5.6 &   7.6 &   9.9 &  11.7 &  11.4 &  10.3 &   9.9 &   8.3 &   7.7 &   8.3 &   9.0 &  10.0 &  11.4 &  12.0 &  13.2 &  14.0 &  15.5 &  15.2 &               9.8 \\
Veterinary        &  18.1 &  17.6 &  16.5 &  16.2 &  15.5 &  16.7 &  13.4 &  20.3 &  21.5 &  22.5 &  20.1 &  19.4 &  21.6 &  21.2 &  20.5 &  22.4 &  24.2 &  24.5 &  25.5 &  26.3 &  27.7 &  27.1 &  29.9 &  27.8 &              21.5 \\
\midrule
All               &  17.6 &  17.8 &  17.7 &  16.9 &  15.6 &  14.7 &  14.9 &  18.7 &  20.5 &  21.8 &  21.7 &  21.8 &  22.5 &  22.5 &  22.7 &  24.4 &  25.8 &  27.0 &  28.1 &  29.4 &  30.3 &  31.1 &  32.5 &  32.4 &              22.9 \\
\bottomrule
\end{tabular}%
}
  \justify \small{\textit{Notes:} The table shows the share of articles with at least one author reporting multiple affiliations. Shares do not account for countries differing shares in fields.}
\end{table}

Looking at the author level, we find that the share of authors with multiple affiliations also increases from 9.8\% in 2001 to 15.9\% in 2019 (see Table \ref{tab:multiaff_authors_share-field}). This corresponds to an increase of more than 60\%. Figure \ref{fig:multiaff_global} graphically reports how author affiliation trends differ by field, journal quality, and country. 
Based on \NOfAuthorfieldyear\ author-field-year observations, multiple affiliations are most common in Neuroscience (20.3\% in 2019) and least common in the Social Sciences (9.3\% in 2019) (Figure \ref{fig:multiaff_global}(A)). Further, relying on \NOfAuthoroctileyear\ author-journal quality group-year observations in Figure \ref{fig:multiaff_global}(B), we see that the number of authors with multiple affiliations increases in all journal quality groups. A journal quality group corresponds to a quartile in the journal impact factor distribution of the journals included in our study (the top 50\%). Top journals have on average a higher share of authors with multiple affiliations.

Figure \ref{fig:multiaff_global}(C) and Figure \ref{fig:multiaff_global}(D) report country differences, where we distinguish the author-year observations by country of first listed affiliation ( \NOfAuthorcountryyear\ observations). Although there are increasing trends in multiple affiliations in most countries, we also see substantial variation. For instance, while in the United States the share of authors with multiple affiliations remains rather constant at about 9\% since 1996, the shares in other larger economies increases substantially over time. For instance, in China the share of authors with multiple affiliations experience an increase already in the late 1990s, peaking in 2005 and remaining relatively constant until 2016 (see Figure \ref{fig:multiaff_global}(C)). In Europe, multiple affiliations become more common relatively abruptly in the early to mid-2000s. France and Norway stand out with a steep increase and a high peak level in 2015, when more than one in four authors report multiple affiliations on their scientific articles (see Figure \ref{fig:multiaff_global}(D)). Indeed, the share of authors with multiple affiliations increases in nearly all countries (see Table \ref{tab:multiaff_authors_share-country} for the full list).

\subsection{Multiple affiliations by organisation type and country}
The increase in multiple affiliations globally could be reflected in increases in affiliations that span different types of organisations (or not) and national borders (or not) (RQ2).

\begin{figure}[ht!]
    \centering
    \caption{Organisation type combinations\label{fig:type_combinations}}
    \includegraphics[width=0.6\textwidth]{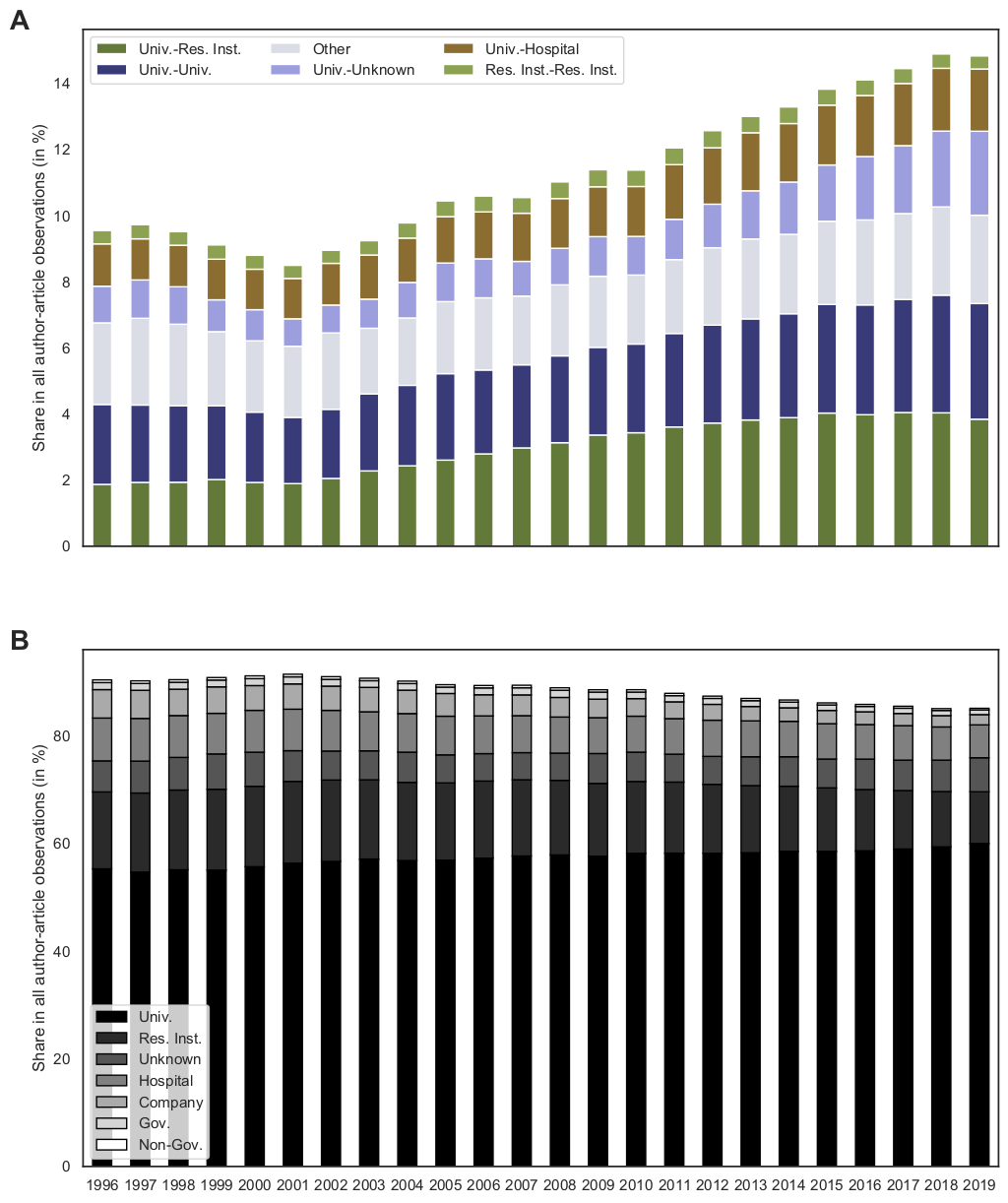}
    \justify \small{\textit{Notes:} Figures depict shares in all affiliations at author-article level. Organisation types are provided by Scopus. Combinations less than 3\% of all combinations in each year grouped as “Other”. (A) Author-article observations with multiple affiliations. (B) Author-article observations with single affiliations. See Table \ref{tab:afftype_share-all_comparison} for precise numbers for both panels.}
\end{figure}

\subsubsection{Organisation types of co-affiliations}

In Figure \ref{fig:type_combinations}, we consider \NOfAuthorpaper\ author-article pairs to investigate organisation type combinations in multiple affiliations. Organisation types (university, research institute, hospital, governmental, non-governmental, private, other) are drawn from Scopus. Panel A shows the different combinations in multiple affiliation. Panel B, for comparison, shows the shares of organisation types based on single affiliation authors, illustrating that university affiliations are by far the most common type. We can see from panel A that about half of all multiple affiliations involve either two universities or a university and a research institute. These are also the combinations that drive the global upward trend in multiple affiliations, with the latter in particular having gained in importance since the early 2000s. This suggests that there may be benefits to both researchers and academic institutions from affiliating across academic institutions. Affiliation with a university and a hospital is very frequently observed, primarily as this represents the most frequent type of co-affiliation in Medicine (26\%; see Table \ref{tab:combs_share-field} for breakdown by field). Co-affiliations that involve governmental or non-governmental organisations (NGO) play only a very minor role. Interestingly, multiple affiliations that involve companies are also relatively rare. This is even true in fields such as Computer Sciences and Engineering. However, not all company or NGO affiliations may be recognized as such, especially if an institution is small and linked affiliations rare and therefore not identified by Scopus. Figure \ref{fig:type_combinations}(A) shows a substantial share of organisation type combinations involving unknown sectors which could be small firms, colleges, government bodies or NGOs.\footnote{A random examination of 100 affiliation profiles with unknown type suggests the following distribution: university and colleges: 40\%, companies: 20\%, hospitals: 12\%, governmental organisations: 9\%, NGOs: 4\%, unclassified: 11\%.} Yet, overall co-affiliations occur largely between academic institutions and less so with organisations outside academia.



\begin{figure}[ht]
    \centering
    \caption{International co-affiliations by country (1996--1999 averages) \label{fig:foreign_affil_shares_19961999}}
    \includegraphics[width=0.6\textwidth]{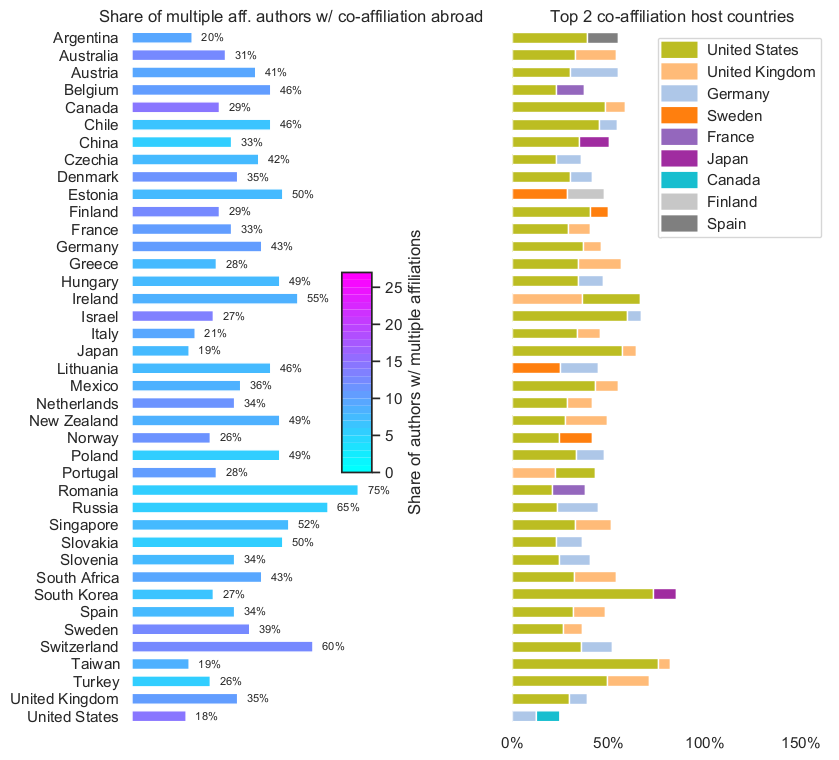}
    \justify \small{\textit{Notes:} The figure depicts the shares of authors with international co-affiliations in all author-article observations with multiple affiliations (left) and the two most common host countries for these affiliations (right) for the 1996-1999 period. Reading example: Argentina has a relatively medium share of authors with multiple affiliations, of which 20\% hold their second affiliation outside Argentina. In about 40\% of cases the second affiliation is in the United States, in a further approx. 10\% it is in Spain.}
\end{figure}

\subsubsection{International co-affiliations}

\begin{figure}[ht]
    \centering
    \caption{International co-affiliation by country (2016--2019 averages) \label{fig:foreign_affil_shares_20162019}}
    \includegraphics[width=0.6\textwidth]{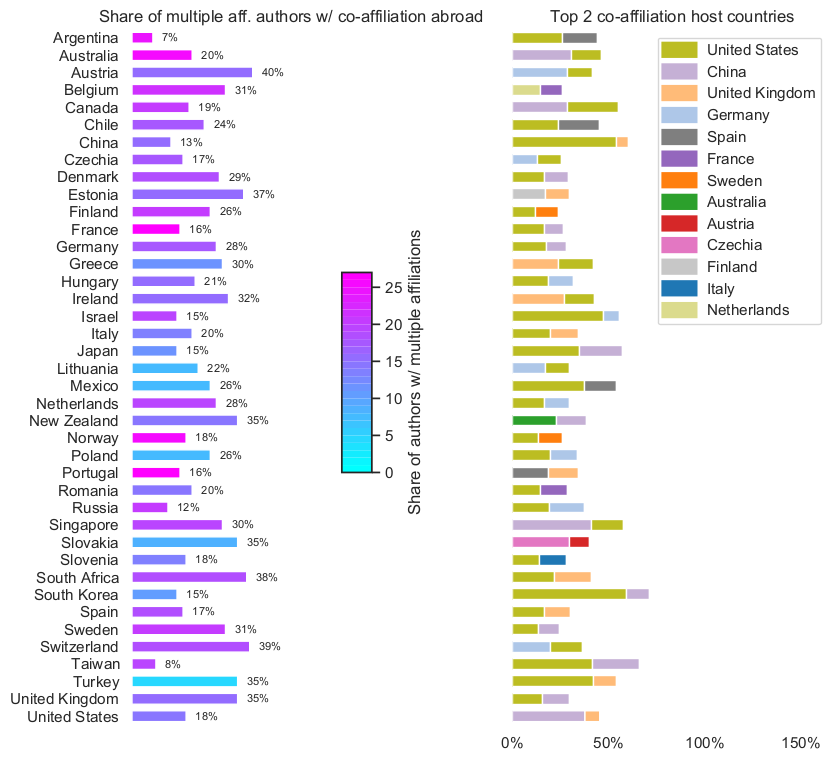}
    \justify \small{\textit{Notes:} The figure depicts the shares of authors with international co-affiliations in all author-article observations with multiple affiliations (left) and the two most common host countries for these affiliations (right) for the 2016-2019 period. Reading example: Argentina has a relatively high share of authors with multiple affiliations, of which 7\% hold their second affiliation outside Argentina. In about 35\% of cases the second affiliation is in the United States, in a further approx. 20\% it is in Spain.}
\end{figure}

Multiple affiliations can moreover occur domestically and internationally. In our data we find that the share of authors with an international co-affiliation in all authors with multiple affiliations differs strongly between countries and the time period considered.\footnote{The share of international co-affiliations also differs by field. See Table \ref{tab:foreignaff_authors_share-countryfield}. The shares are highest in Economics (55.2\% on average over all years), Business (51.5\%) and Decision Sciences (48.3\%), and lowest in Nursing (19.9\% and Health (24.2\%).} The left-hand panels in Figures \ref{fig:foreign_affil_shares_19961999} and \ref{fig:foreign_affil_shares_20162019} show this share for all 40 countries during the respective time periods. Colour shades indicate the share of authors with multiple affiliations in all authors in each country in the same period.
Comparing the two we can see that the overall importance of international co-affiliations decreases over time. In the 1996--1999 period the share of authors with an international co-affiliation in all authors with multiple affiliations ranges between 18\% for the United States and 75\% for Romania. In the 2016-2019 the shares range between 7\% in Argentina and 40\% in Austria, while the overall proportion of multiple affiliations increased. Indeed, the country-level correlation between the share of authors with multiple affiliations and the share of authors with an international co-affiliation is $\rho = -0.295$ (1996--1999: $\rho = -0.380$), indicating that in countries where multiple affiliations are more frequent, these mainly occur domestically.

Figures \ref{fig:foreign_affil_shares_19961999} and \ref{fig:foreign_affil_shares_20162019} moreover illustrate the importance of a small number of "host" countries, depicted in the right-hand panels. 
For the majority of countries the most important "host" of a co-affiliation is the United States. Frequent countries among the two most important ones are China, United Kingdom and Germany. China is also the most important "host" for international affiliations of United States-based researchers. In some instances the most important "host" is a neighboring country, e.g. the Netherlands for Belgium or Spain for Portugal (right part of Figure \ref{fig:foreign_affil_shares_20162019}). Comparing the 2016--2019 period to the 1996--1999 period reveals a broadened set of host countries as China and several European countries became more prominent. Moreover, we see an overall decrease in the concentration in top partner countries: The United States and the United Kingdom remain very prominent, but several European countries, such as Germany and Spain, and particularly China emerge amongst the top 2 hosts of international affiliations in more recent years. 

\begin{figure}[ht]
    \centering
    \caption{Leiden communities of countries based on international co-affiliation (2016-2019 averages) \label{fig:cluster_20162019}}
    \includegraphics[width=0.6\textwidth]{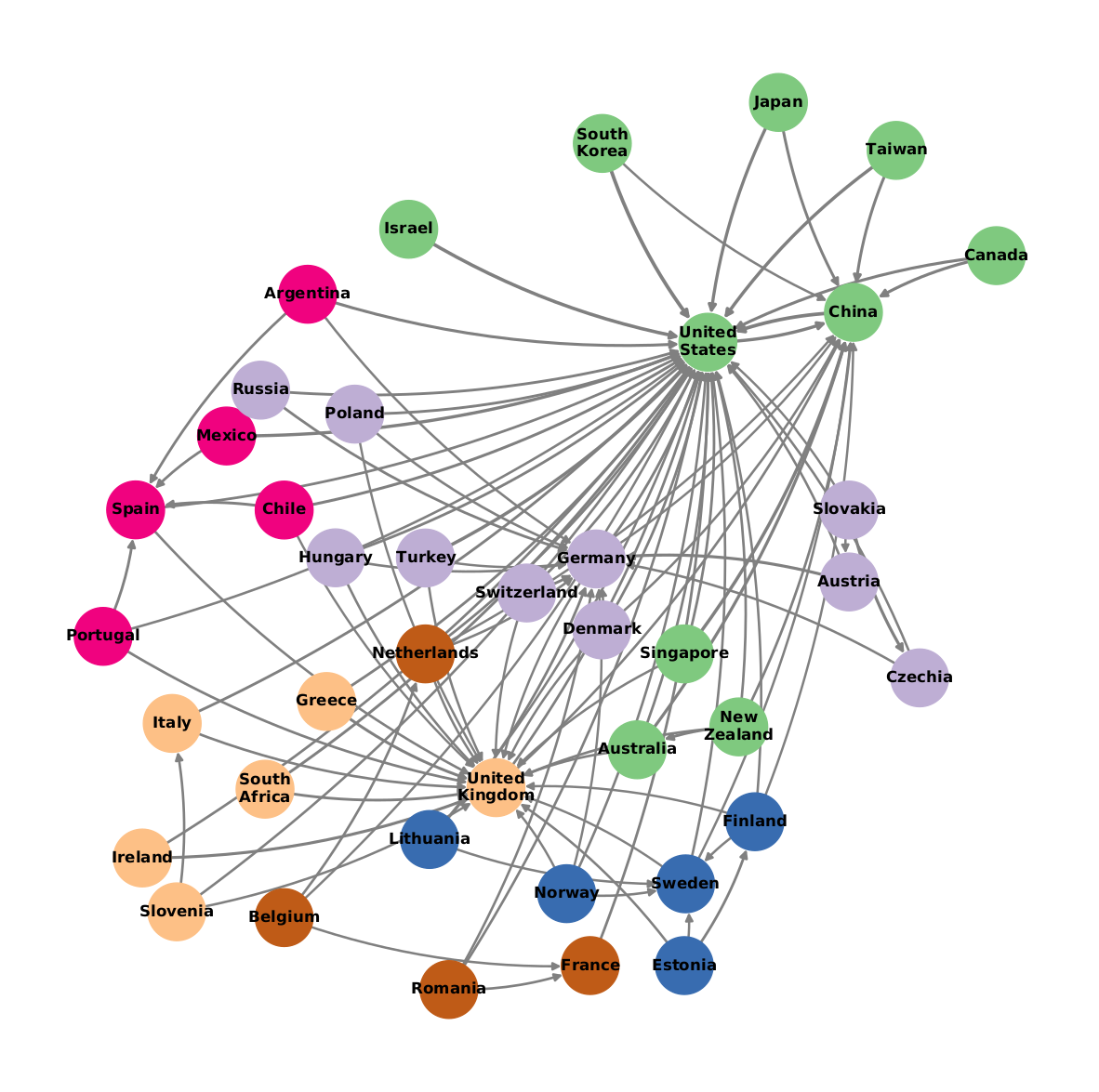}
    \justify \small{\textit{Notes:} The figure depicts communities of countries based on international co-affiliations of authors from a source (home) country. A source node is connected to another if a share not less than 10\% of authors with international co-affiliation from the source hold their co-affiliation in the other country. Node colors indicate joint community based on the Leiden algorithm. Link width is proportional to the share of authors linked with the target country. Graph analysis conducted with code provided by \citet{Hagberg2004NetworkX} and \citet{RosettiMiliCazabet2019}.}
\end{figure}

There are also groups of countries that tend to maintain higher levels of international co-affiliation among themselves rather than with others. Figure \ref{fig:cluster_20162019} reveals such communities according to the Leiden algorithm \citep{TraagWaltmanvanEck2019} based on host linkages when the link represents at least 10\% of country authors with an international co-affiliation. 

While almost all countries are linked to the United States, there are countries additionally (or more importantly) linked to one common country. For example, Argentina, Chile, Mexico and Portugal are additionally linked to Spain. Sweden and France emerge as centres of small communities, while Germany is the centre of a larger central European community. Interestingly China and the United States are closely interlinked as part of a largely Pacific community. While all these communities share clear geographic or language linkages, a more diverse community emerges with the United Kingdom at its centre, comprising Greece, Ireland, Italy, Slovenia, and South Africa. Here we may observe the result of longer term brain drain to the United Kingdom \citep{vanderwende_2015}. 

Figure \ref{fig:cluster_20162019} also hints at a core-periphery structure in the network, where few countries are common destinations for international co-affiliations. This is further apparent from Figures \ref{fig:foreign_affil_network_19961999} and \ref{fig:foreign_affil_network_20162019} which depict the information from Figures \ref{fig:foreign_affil_shares_19961999} and \ref{fig:foreign_affil_shares_20162019} in a network perspective, corroborating the central position of especially the United States, China, the United Kingdom and Germany in the 2016--2019 period. Other community centres, such as France and the Netherlands are rather peripheral in such a network. 

\FloatBarrier

\subsection{The role of Excellence Initiatives}
The relatively sudden rise in the prevalence of multiple affiliations in the 2000s and the observed country differences may be related to reforms in the allocation of research funding that several countries have experienced during that period (RQ3). 

\begin{figure}[htb]
    \centering
    \caption{Share of authors with multiple affiliations averaged over countries by year \label{fig:multiaff_groups-countryfield}}
    \includegraphics[width=0.6\textwidth]{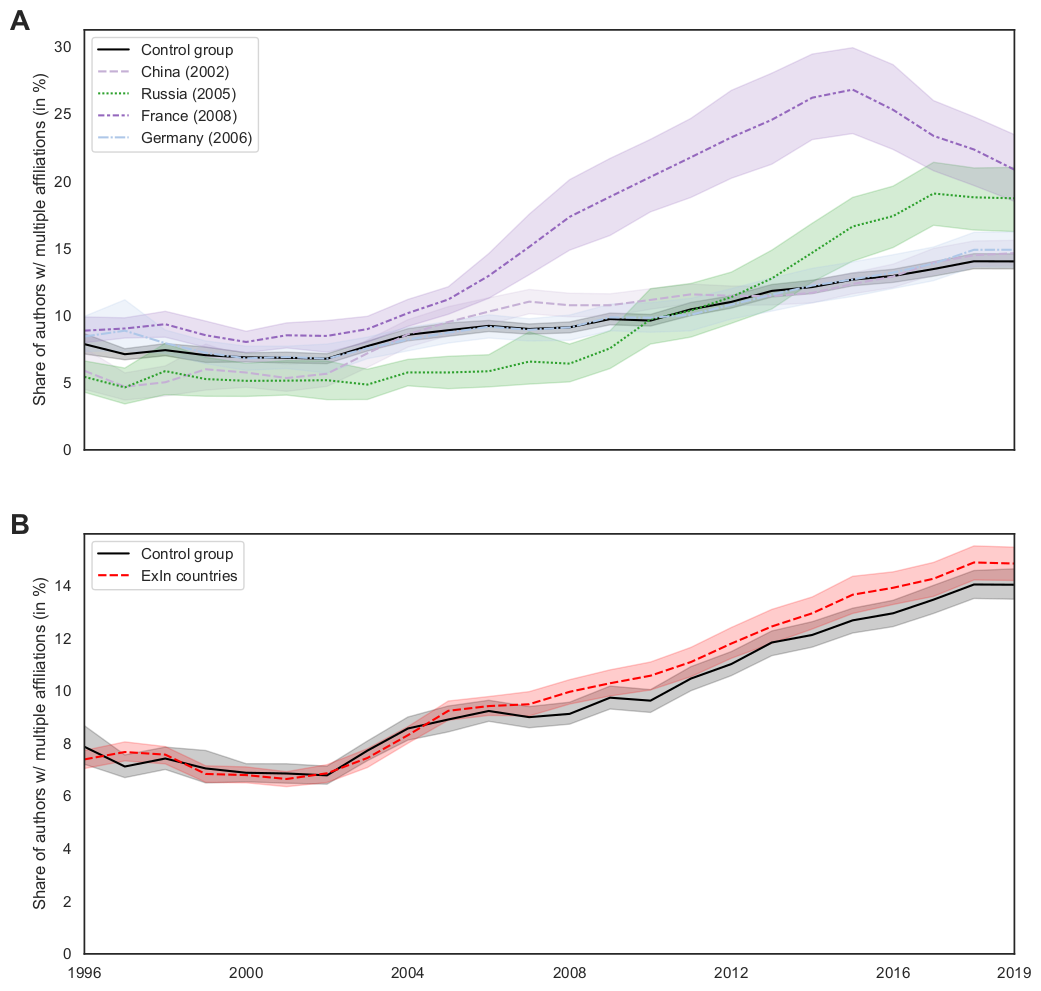}
    \justify \small{\textit{Notes:} The figure shows the share of authors with multiple affiliations accounting for countries' differing share in fields, $\overline{s}_{c,t}$. Shaded area indicates 95\% confidence interval. Plot uses \NOfAuthorcountryfieldyear\ author-field-country-year observations. See Tables \ref{tab:multiaff_authors_share-fieldcountry} and \ref{tab:multiaff_groups-countryfield} for precise values.}
\end{figure}

We collected information on ExIns from \cite{Froumin2015} and \cite{GEUNA2016260}. Some countries introduced several initiatives during the period of our study. For the purpose of our analysis we focus on the first introduction and only consider initiatives that involved substantial financial resources, i.e. more than US\$1 million. We identified 17 countries that introduced ExIns between 2002 and 2018. Figure \ref{fig:multiaff_global}(C/D) above showed that before 2002 most countries had seen constant (or even slightly declining) shares of authors with multiple affiliations. Looking at Figure \ref{fig:multiaff_groups-countryfield}, which shows the trend in the share of multiple affiliations for countries that did not introduce ExIns (control group) and those that did (ExIn), we can indeed see how closely the trends correspond in those early years. The share of multiple affiliations started to increase in the early 2000s in countries with and without ExIns (see also Figure \ref{fig:multiaff_countriesmatrix-countryfield}), though this increase appears stronger for the ExIn group. Figure \ref{fig:multiaff_groups-countryfield}(A) reports trends for selected ExIn countries and indicates some heterogeneity amongst these, with Germany staying close to the control group, while the Russia experiences an increase some years after the first ExIn. 

The question arises whether the increase in multiple affiliations we can observe descriptively is indeed significantly stronger after the introduction of ExIns. To investigate this, we conduct difference-in-differences (DiD) analyses. In particular, we compare the likelihood that an author located in a country that introduced some kind of reform has multiple affiliations to the likelihood in a pool of authors located in control countries that did not significantly change their science funding structures. The DiD is the difference in this comparison before and after the year the reform came into effect. 

We perform separate regressions at the author-article level for each ExIn country, as ExIns were introduced at different points in time. We hold the control group of countries constant\footnote{Countries in the control group are Austria, Belgium, Chile, Czech Republic, Estonia, Finland, Greece, Hungary, Ireland, Lithuania, Mexico, The Netherlands, New Zealand, Portugal, Romania, Slovakia, Slovenia, South Africa, Sweden, Switzerland, Turkey and the United States of America.}, i.e. we compare each ExIn country to the same set of control countries which corresponds to the control group in Figure \ref{fig:multiaff_groups-countryfield}. We estimate linear probability models (LPM) for the likelihood that an author has more than one affiliation listed on an article, and include the number of co-authors, and the time-varying Scimago Journal Rank as control variables. We also include journal, country, and year fixed effects. 

Table \ref{tab:ols_exin} shows the results from the different country DiD models, reporting the year of the respective ExIns in the top row. Note that the unit of observation is article-author combination to account for the possibility of international co-authorship (i.e. an article to be assigned to more than one country) and multiple affiliations of several co-authors on an article. When interpreting the results, we need to note that there are likely announcement effects at play, which may result in behavioural changes prior the formal starting point of the reforms. Since the duration of the post-reform period is naturally longer for countries that introduced reforms earlier, we estimate models for the maximum post period (Table \ref{tab:ols_exin}, and for a fixed window of one decade for those countries that implemented reforms prior to 2008 (Table \ref{tab:ols_exin-short}).      

Our analysis reveals a significant increase in multiple affiliations after the introduction of funding reforms in most of the 17 ExIn cases, where the results (for the full time window) show a positive and significant DiD (ExIn $\times$ Post-ExIn). The DiD estimates are largest in the case of France and Norway with an increase in the likelihood to have an author with multiple affiliations that is 11 percentage points higher than in the control group. Russia shows a higher likelihood of 9 percentage points compared to the control group. Australia, Singapore, Taiwan and Spain show also significantly higher likelihoods in the occurrence of co-affiliated authors. Smaller differences can be found in Canada, China, Denmark, Israel, and Germany with around two to three percentage points higher increases than the control group. Comparing these results to those with a fixed time window (Table \ref{tab:ols_exin-short}) shows that for Japan, the effect is larger in the short-run than in the long-run (where it is insignificant) unlike in other countries where multiple affiliations continue to increase. 
For the United Kingdom and Italy that have both undertaken similar research assessments of universities in 2008 \citep{GEUNA2016260}, we do find a positive, but not statistically significant increase in the likelihood to see authors with multiple affiliations. This suggest that selective funding mechanisms and not research evaluation systems as such may be associated with multiple affiliations. We also do not find indications for treatment effects for South Korea. A negative DiD can be observed for Poland that - as the only ExIn country - sees significantly fewer authors with multiple affiliations after the implementation of ExIns. 
The control variables show that the number of authors on an article is negatively associated with the likelihood of seeing an author with multiple affiliations indicating that co-affiliations may also serve as a substitute for co-authorship.

\begin{landscape}
\begin{table}[b]
  \caption{Difference-in-differences estimation (linear probability models) on the occurrence of a multiple affiliation author on an article\label{tab:ols_exin}}
  \begin{center}
      {\scriptsize \begin{tabular}{@{\extracolsep{0pt}}ld{2.3}d{2.3}d{2.3}d{2.3}d{2.3}d{2.3}d{2.3}d{2.3}d{2.3}}
\toprule
Country & \multicolumn{1}{c}{\textbf{Japan}} & \multicolumn{1}{c}{\textbf{China}} & \multicolumn{1}{c}{\textbf{Norway}} & \multicolumn{1}{c}{\textbf{Australia}} & \multicolumn{1}{c}{\textbf{South Korea}} & \multicolumn{1}{c}{\textbf{Russia}} & \multicolumn{1}{c}{\textbf{Germany}} & \multicolumn{1}{c}{\textbf{Singapore}} \\
Year of ExIn & \multicolumn{1}{c}{2002} & \multicolumn{1}{c}{2002} & \multicolumn{1}{c}{2003} & \multicolumn{1}{c}{2003} & \multicolumn{1}{c}{2004} & \multicolumn{1}{c}{2005} & \multicolumn{1}{c}{2006} & \multicolumn{1}{c}{2006} &\\
\midrule
ExIn              & -0.088^{***}  & -0.072^{***}  & -0.067^{***}& -0.061^{***} & -0.091^{***} & -0.082^{***} & -0.082^{***} & -0.070^{***} &   \\
                  & (0.008)       & (0.006)       & (0.007)     & (0.007)      & (0.006)      & (0.005)      & (0.007)      & (0.005)     & \\
Post-ExIn         & 0.042^{**}    & 0.046^{**}    & 0.041^{**}  & 0.043^{**}   & 0.040^{**}   & 0.042^{**}   & 0.043^{**}   & 0.041^{**} & \\
                  & (0.016)       & (0.017)       & (0.017)     & (0.018)      & (0.016)      & (0.017)      & (0.017)      & (0.017)    & \\
ExIn x Post-ExIn  & 0.004         & 0.027^{***}   & 0.113^{***} & 0.067^{***}  & -0.006       & 0.092^{***}  & 0.025^{***}  & 0.056^{***} & \\
                  & (0.008)       & (0.008)       & (0.009)     & (0.009)      & (0.007)      & (0.009)      & (0.009)      & (0.008)     & \\
\# of authors     & -0.013^{***}  & -0.014^{***}  & -0.013^{***}& -0.012^{***} & -0.012^{***} & -0.013^{***} & -0.013^{***} & -0.013^{***}  & \\
                  & (0.001)       & (0.002)       & (0.001)     & (0.001)      & (0.001)      & (0.001)      & (0.001)      & (0.001)      & \\
Journal Rank      & 0.001         & 0.001         & 0.001       & 0.001        & 0.001        & 0.001        & 0.001        & 0.001 & \\
                  & (0.001)       & (0.001)       & (0.001)     & (0.001)      & (0.001)      & (0.001)     & (0.001)       & (0.001)  &  \\
                  \midrule
Journal-fixed effects & \multicolumn{1}{c}{\checkmark} & \multicolumn{1}{c}{\checkmark} & \multicolumn{1}{c}{\checkmark} & \multicolumn{1}{c}{\checkmark} & \multicolumn{1}{c}{\checkmark} & \multicolumn{1}{c}{\checkmark} & \multicolumn{1}{c}{\checkmark} & \multicolumn{1}{c}{\checkmark} & \\
Country-fixed effects & \multicolumn{1}{c}{\checkmark} & \multicolumn{1}{c}{\checkmark} & \multicolumn{1}{c}{\checkmark} & \multicolumn{1}{c}{\checkmark} & \multicolumn{1}{c}{\checkmark} & \multicolumn{1}{c}{\checkmark} & \multicolumn{1}{c}{\checkmark} & \multicolumn{1}{c}{\checkmark} & \\
Year-fixed effects & \multicolumn{1}{c}{\checkmark} & \multicolumn{1}{c}{\checkmark} & \multicolumn{1}{c}{\checkmark} & \multicolumn{1}{c}{\checkmark} & \multicolumn{1}{c}{\checkmark} & \multicolumn{1}{c}{\checkmark} & \multicolumn{1}{c}{\checkmark} & \multicolumn{1}{c}{\checkmark} & \\
$R^{2}$  & \multicolumn{1}{c}{0.026} & \multicolumn{1}{c}{0.022} & \multicolumn{1}{c}{0.027} & \multicolumn{1}{c}{0.029} & \multicolumn{1}{c}{0.026} & \multicolumn{1}{c}{0.026} & \multicolumn{1}{c}{0.025} & \multicolumn{1}{c}{0.025} & \\
\# observations & \multicolumn{1}{c}{46,447,230} & \multicolumn{1}{c}{53,151,993} & \multicolumn{1}{c}{39,735,402} & \multicolumn{1}{c}{41,795,377} & \multicolumn{1}{c}{42,174,198} & \multicolumn{1}{c}{40,224,951} & \multicolumn{1}{c}{45,014,574} & \multicolumn{1}{c}{39,707,842} & \\
\bottomrule
&       &       &       &       &       &       &       &  & \\
\ctoprule{1-10}
Country  & \multicolumn{1}{c}{\textbf{Taiwan}} & \multicolumn{1}{c}{\textbf{Denmark}} & \multicolumn{1}{c}{\textbf{France}} & \multicolumn{1}{c}{\textbf{UK}} & \multicolumn{1}{c}{\textbf{Italy}} & \multicolumn{1}{c}{\textbf{Spain}} & \multicolumn{1}{c}{\textbf{Israel}} & \multicolumn{1}{c}{\textbf{Poland}}  &  \multicolumn{1}{c}{\textbf{Canada}}  \\
Year of ExIn  & \multicolumn{1}{c}{2006} & \multicolumn{1}{c}{2008} & \multicolumn{1}{c}{2008} & \multicolumn{1}{c}{2008} & \multicolumn{1}{c}{2008} & \multicolumn{1}{c}{2009} & \multicolumn{1}{c}{2010} & \multicolumn{1}{c}{2012} & \multicolumn{1}{c}{2014}  \\
\cmidrule{1-10}
ExIn             & -0.066^{***} & -0.073^{***} & -0.056^{***} & -0.081^{***}& -0.078^{***}& -0.085^{***} & -0.046^{***}& -0.089^{***} &  -0.032^{***}  \\
                 & (0.007)      & (0.006)      & (0.006)      & (0.006)     & (0.006)     & (0.005)      & (0.006)     & (0.004)      &   (0.002)        \\
Post-ExIn        & 0.041^{**}   & 0.041^{**}   & 0.041^{**}   & 0.041^{**}  & 0.023^{**}  & 0.043^{**}   & 0.041^{**}  & 0.040^{**}  &   0.040^{**}   \\
                 & (0.017)      & (0.016)      & (0.016)      & (0.016)     & (0.009)     & (0.018)      & (0.017)     & (0.016)      &   (0.016)   \\
ExIn x Post-ExIn & 0.065^{***}  & 0.029^{***}  & 0.112^{***}  & 0.011       & 0.011       & 0.068^{***}  & 0.020^{**}  & -0.018^{**} &  0.018^{***}       \\ 
                 & (0.008)      & (0.009)      & (0.009)      & (0.009)     & (0.009)     & (0.009)      & (0.009)     & (0.008)      &  (0.008)         \\
\# of authors    & -0.012^{***} & -0.013^{***} & -0.014^{***} & -0.012^{***}&-0.011^{***} & -0.012^{***} & -0.012^{***}& -0.013^{***}  &  -0.012^{***}       \\
                 & (0.001)      & (0.001)      & (0.002)      & (0.001)     & (0.009)     & (0.001)      & (0.001)     & (0.001)      &  0.001       \\
Journal Rank     & 0.001        & 0.001        & 0.001        & 0.001       & 0.001       & 0.002        & 0.001       & 0.001        & 0.001         \\
                 & (0.001)      & (0.001)      & (0.001)      & (0.001)     & (0.001)     & (0.001)      & (0.001)     & (0.001)      &  (0.001)         \\
\cmidrule{1-10}
Journal-fixed effects & \multicolumn{1}{c}{\checkmark} & \multicolumn{1}{c}{\checkmark} & \multicolumn{1}{c}{\checkmark} & \multicolumn{1}{c}{\checkmark} & \multicolumn{1}{c}{\checkmark} & \multicolumn{1}{c}{\checkmark} & \multicolumn{1}{c}{\checkmark} & \multicolumn{1}{c}{\checkmark}  & \multicolumn{1}{c}{\checkmark} \\
Year-fixed effects & \multicolumn{1}{c}{\checkmark} & \multicolumn{1}{c}{\checkmark} & \multicolumn{1}{c}{\checkmark} & \multicolumn{1}{c}{\checkmark} & \multicolumn{1}{c}{\checkmark} & \multicolumn{1}{c}{\checkmark} & \multicolumn{1}{c}{\checkmark} & \multicolumn{1}{c}{\checkmark}  & \multicolumn{1}{c}{\checkmark} \\
$R^{2}$ & \multicolumn{1}{c}{0.027} & \multicolumn{1}{c}{0.025} & \multicolumn{1}{c}{0.034} & \multicolumn{1}{c}{0.025} & \multicolumn{1}{c}{0.024} & \multicolumn{1}{c}{0.028} & \multicolumn{1}{c}{0.025} & \multicolumn{1}{c}{0.025} & \multicolumn{1}{c}{0.027}  \\
\# observations & \multicolumn{1}{c}{40,666,505} & \multicolumn{1}{c}{39,983,677} & \multicolumn{1}{c}{43,536,921} & \multicolumn{1}{c}{45,214,245} & \multicolumn{1}{c}{43,791,023} & \multicolumn{1}{c}{42,227,576} & \multicolumn{1}{c}{39,941,934} & \multicolumn{1}{c}{40,107,934}  & \multicolumn{1}{c}{42,483,035}\\
\cbottomrule{1-10}
\end{tabular}
}
  \end{center}
  \justify \small{\textit{Notes:} The table shows estimated coefficients and country-clustered standard errors in parenthesis. All models contain journal fixed effects a constant. *** (**, *) indicate a 1\% (5\%, 10\%) significance level. The control group consists of Austria,  Belgium,  Chile,  Czech Republic,  Estonia,  Finland,  Greece,  Hungary, Ireland, Lithuania, Mexico, The Netherlands, New Zealand, Portugal, Romania, Slovakia, Slovenia, South Africa, Sweden, Switzerland, Turkey, and the United States of America accounting for 39,211,604 author-article observations.}
 
\end{table}
\end{landscape}
\FloatBarrier

\begin{landscape}
\begin{table}[b]
  \centering
  \caption{Difference-in-differences estimation (linear probability models) on the occurrence of a multiple affiliation author on an article with fixed post-initiative window\label{tab:ols_exin-short}}
  {\scriptsize \begin{tabular}{@{\extracolsep{0pt}}ld{2.1}d{2.1}d{2.1}d{2.1}d{2.1}d{2.1}d{2.1}d{2.1}d{2.1}}
\toprule
Country & \multicolumn{1}{c}{\textbf{Japan}} & \multicolumn{1}{c}{\textbf{China}} & \multicolumn{1}{c}{\textbf{Norway}} & \multicolumn{1}{c}{\textbf{Australia}} & \multicolumn{1}{c}{\textbf{South Korea}} & \multicolumn{1}{c}{\textbf{Russia}} & \multicolumn{1}{c}{\textbf{Germany}} & \multicolumn{1}{c}{\textbf{Singapore}} & \multicolumn{1}{c}{\textbf{Taiwan}} \\
Year of ExIn & \multicolumn{1}{c}{2002} & \multicolumn{1}{c}{2002} & \multicolumn{1}{c}{2003} & \multicolumn{1}{c}{2003} & \multicolumn{1}{c}{2004} & \multicolumn{1}{c}{2005} & \multicolumn{1}{c}{2006} & \multicolumn{1}{c}{2006} & \multicolumn{1}{c}{2006} \\
\midrule
ExIn             & -0.062^{***}  & -0.045^{***}  & -0.045^{***}  & -0.038^{***} & -0.073^{***}  & -0.072^{***} & -0.069^{***} & -0.059^{***}  &  -0.054^{***}  \\
                 & (0.004)       & (0.002)       & (0.004)       & (0.004)      & (0.003)       & (0.002)      & (0.004)      & (0.003)       &  (0.005)      \\
Post-ExIn        & 0.017^{*}      & 0.015^{*}    & 0.020^{*}     & 0.021^{*}    & 0.023^{**}    & 0.027^{**}   & 0.031^{**}   & 0.030^{**}    &  0.030^{**}    \\
                 & (0.070)       & (0.009)       & (0.011)       & (0.011)      & (0.011)       & (0.012)      & (0.012)      & (0.012)       &  (0.013)      \\
ExIn x Post-ExIn & 0.008^{*}     & 0.026^{***}   & 0.086^{***}   & 0.044^{***}  & -0.006        & 0.048^{***}  & 0.016^{**}   & 0.046^{***}   &  0.054^{***}  \\
                 & (0.004)       & (0.004)       & (0.006)       & (0.005)      & (0.005)       & (0.006)      & (0.007)      & (0.006)       &  (0.007)      \\
ln(\# of authors)& -0.014^{***}  & -0.015^{***}   & -0.013^{***} & -0.013^{***}  & -0.013^{***} & -0.013^{***} & -0.013^{***} & -0.012^{***}  &  -0.012^{***}  \\
                 & (0.001)       & (0.001)       & (0.001)       & (0.001)      & (0.001)       & (0.001)      & (0.001)      & (0.001)       &  (0.001)      \\
Journal Rank     & -0.001^{***}  & -0.001^{**}    & -0.001       & -0.001      & -0.001        & -0.001       & 0.001        & 0.001         &   0.001     \\
                 & (0.001)       & (0.001)       & (0.001)       & (0.001)      & (0.001)       & (0.001)      & (0.001)      & (0.001)       &   (0.001)     \\
\midrule
Journal-fixed effects & \multicolumn{1}{c}{\checkmark} & \multicolumn{1}{c}{\checkmark} & \multicolumn{1}{c}{\checkmark} & \multicolumn{1}{c}{\checkmark} & \multicolumn{1}{c}{\checkmark} & \multicolumn{1}{c}{\checkmark} & \multicolumn{1}{c}{\checkmark} & \multicolumn{1}{c}{\checkmark}  & \multicolumn{1}{c}{\checkmark}\\
Year-fixed effects & \multicolumn{1}{c}{\checkmark} & \multicolumn{1}{c}{\checkmark} & \multicolumn{1}{c}{\checkmark} & \multicolumn{1}{c}{\checkmark} & \multicolumn{1}{c}{\checkmark} & \multicolumn{1}{c}{\checkmark} & \multicolumn{1}{c}{\checkmark} & \multicolumn{1}{c}{\checkmark} & \multicolumn{1}{c}{\checkmark}\\
 $R^{2}$ & \multicolumn{1}{c}{0.025} & \multicolumn{1}{c}{0.023} & \multicolumn{1}{c}{0.025} & \multicolumn{1}{c}{0.025} & \multicolumn{1}{c}{0.025} & \multicolumn{1}{c}{0.024} & \multicolumn{1}{c}{0.023} & \multicolumn{1}{c}{0.024} & \multicolumn{1}{c}{0.025} \\
\# observations & \multicolumn{1}{c}{24,963,993} & \multicolumn{1}{c}{23,861,792} & \multicolumn{1}{c}{22,855,785} & \multicolumn{1}{c}{23,830,670} & \multicolumn{1}{c}{26,375,022} & \multicolumn{1}{c}{27,684,665} & \multicolumn{1}{c}{33,713,562} & \multicolumn{1}{c}{29,698,139} & \multicolumn{1}{c}{30,463,787} \\
\cbottomrule{1-10} 
\end{tabular}
  }
  \justify \small{\textit{Notes:} The table shows estimated coefficients and standard errors in parenthesis. All models contain a constant *** (**, *) indicate a 1\% (5\%, 10\%) significance level. The control group consists of Austria,  Belgium,  Chile,  Czech Republic,  Estonia,  Finland,  Greece,  Hungary, Ireland, Lithuania, Mexico, The Netherlands, New Zealand, Portugal, Romania, Slovakia, Slovenia, South Africa, Sweden, Switzerland, Turkey, and the United States of America.}
\end{table}
\end{landscape}
\FloatBarrier

\section{Conclusions}
This paper analysed affiliation information on published articles in 40 countries and 26 scientific fields over a 24 year period. We set out to answer three research questions: (RQ1) how multiple affiliations evolved over time in different countries and fields, (RQ2) whether these changes are corresponding to within or across sectors, and to domestic or international co-affiliations, and (RQ3) how they are impacted by ExIns. 

Our findings regarding RQ1, reported in Section 3.1, showed substantial increases in multiple affiliations across countries and fields. We showed that they are prevalent in all countries, with substantial increases in multiple affiliations in France and Russia amongst others, and no increase in the United States.  
The increasing prevalence of multiple affiliations suggests that fundamental changes to institutional conditions and the organization of science are at work. Previous research discussed changes in the complexity of science and increases in team sizes and cross-institutional collaborations on co-authored papers as its consequence and important coping mechanisms \citep{ADAMS2005, Wuchty1036, Jones2008}. The rise in multiple affiliations may be the reflection of another coping mechanism. 

In response to RQ2, we documented different types of multiple affiliation in terms of the involved organisations and countries, reported in Section 3.2. We found that the majority of co-affiliations are between academic institutions, who drive the upwards trend. Co-affiliations between PROs and universities, in particular, have seen an increase. We further found that international co-affiliations account for slightly more than a fifth of all multiple affiliations and, while global network patterns of these international linkages changed over time, their relative importance did not increase. This indicates that the rise in multiple affiliations cannot be explained by a growth in international co-affiliations alone. Indeed, countries with higher increases have seen mainly more domestic co-affiliations.

Domestic science and higher education policies could be a critical driver behind these changes. To answer RQ3 we therefore investigated the effect of ExIns on multiple affiliations in difference-in-difference estimations, reported in Section 3.3. We show that the increase of authors with multiple affiliations has been particularly pronounced in countries that implemented substantial structural funding reforms over the past two decades. Examples include China (2002), Norway (2003), Russia (2005), Germany (2006), Singapore (2008), France (2008) and Israel (2010) \citep{Salmi2916,Schiermeier2017, Butler2009}. A shift in national research funding towards a higher concentration of resources and research output in fewer (elite) places \citep{Hamann2018,Schier2015} may constitute an incentive to affiliate with multiple institutions. In particular, co-affiliations to well-endowed institutions could provide a means for individual researchers to redress any imbalances in resource access. Some of the funding allocation mechanisms explicitly encourage collaboration between organisations which may result in authors listing all involved organisations on their publications as to share the output recognition between all of these. 
In addition, reorganisation in science to gain visibility in international rankings has also been observed, and in the case of France can directly explain the increase in domestic co-affiliations following funding reforms \citep{Paradeise2018}.

While our analysis provides some new insights, there are other potential mechanisms we need to acknowledge. International co-affiliations could reflect traces of increased international mobility \citep{Schiermeier2017, krieger2016} and may provide an important source of productivity-enhancing "home country linkages" \citep{Baruffaldi2012}, or make it easier for researchers to stay connected with previous institutions when internationally mobile. We find evidence of such traces in the co-affiliation communities, in particular the community centred around the United Kingdom. Such co-affiliations are likely beneficial for international research networks and knowledge exchange, contributing to brain circulation \citep{LANGA2018}. Yet, the increase in multiple affiliations is largely driven by domestic, not international co-affiliations, and international mobility likely a smaller contributing factor. An increase in international affiliations is generally seen in smaller science systems with perhaps weaker academic institutions. The United Kingdom and Switzerland are the two leading science nations with an above average share, which possibly reflects their status as a destination for foreign scholars and embeddedness in diaspora networks.

The increase in multiple affiliations may also be related to the growing importance of bibliometric indicators for research funding distribution more generally \citep{Geuna1997}. Institutions may have strong incentives to affiliate prolific researchers in order to increase their chances in funding competitions and to improve their ranking in institutional assessments. We indeed find a higher share of authors with multiple affiliations on articles in top journals compared to lower impact journals, which is indicative of such ranking mechanisms. 

In terms of policy implications, blanket condemnation of multiple affiliations is not a good fix, as researchers and institutions can benefit. Yet, the increase in multiple affiliations also implies that counting publications simply based on listed affiliations distorts institutional performance measures and rankings, leading to difficulties in assigning research efforts and investment to individual universities. This adds to existing concerns about research metrics currently used to inform science policy \citep{deRijcke2015}. Multiple affiliations should therefore be taken into account in any bibliometric evaluations. Just as we acknowledge the contribution that multiple authors make to a scientific discovery, we may explicitly acknowledge the contribution of multiple institutions to the scientific work of an author. It may, however, require consensus about when the listing of an affiliation is justified based on its contribution. 

Multiple affiliations may also reflect (or may be a symptom of) a decline of institutional support for academics, especially regarding resource constraints in university based research or the casualisation of the academic profession, which require academics to seek resources and work roles outside their main institution. These consequences of resource concentration and stratification in higher education need to be considered by science funders. 

Our analysis is, however, subjected to some limitations. 
One limitation arises from heterogeneous standards on how affiliations are reported on a paper, which may affect the assignment of authors to affiliations, leading to over- or under-counting of affiliations. 

Others arise from some of our assumptions. First, we assume that in case of multiple affiliations, the first listed affiliation is the main affiliation of an author. This might not always be the case, and consequently we may assign the wrong host country. The assignment of authors to countries also impacts the regression-based analysis which may result in an underestimation of the observed increases in countries with funding reforms due to authors being assigned to other countries based on first listing. Second, we assume that multiple affiliations on publications are co-occurring, when instead they may represent author mobility, with authors listing both their old and new employer to recognise the contribution of both. These limitations are inherent in bibliometric data and multiple affiliations therefore warrant further investigation using other types of data.

Finally, while we investigated ExIns as a potential driver of multiple affiliations, there may be other important factors driving these developments. We therefore encourage more research on this topic.

\section{Instructions for the replication of the analysis}
The programming code files to replicate the data collection, all calculations and the data analysis are available at https://github.com/Michael-E-Rose/The-Rise-of-Multiple-Institutional-Affiliations. We used the Python packages pybliometrics to retrieve the data \citep{Rose2019}, pandas to manipulate the data \citep{McKinney2010DataPython}, seaborn for visualisation \citep{Waskom2018Mwaskom/seaborn:2018}, and NetworkX for network analysis \citep{Hagberg2004NetworkX}. Note that replication requires subscription to Scopus. If random sampling was applied, the seed was always equal to zero.

\theendnotes

\newpage
\addcontentsline{toc}{section}{References}
\begin{singlespace}
\bibliographystyle{apa}
\bibliography{Meta/references}
\end{singlespace}

\newpage
\appendix
\pagenumbering{Roman}
\setcounter{page}{1}
\setcounter{table}{0}
\renewcommand{\thetable}{A\arabic{table}}
\setcounter{figure}{0}
\renewcommand{\thefigure}{A\arabic{figure}}
\section{Appendix: Additional tables and graphs}

\begin{table}[H]
 \caption{Number of authors and articles by country of first affiliation \label{tab:articlesauthors_country}}
 \centering
 {\scriptsize \begin{tabular}{lrrrr}
\toprule
{} & \multicolumn{2}{c}{Articles} & \multicolumn{2}{c}{Authors} \\
{} &    Unique & Share (in \%) &    Unique & Share (in \%) \\
\midrule
Argentina      &   113,862 &         0.46 &    66,346 &         0.45 \\
Australia      &   785,613 &         3.14 &   340,790 &         2.31 \\
Austria        &   194,735 &         0.78 &    91,460 &         0.62 \\
Belgium        &   307,585 &         1.23 &   133,103 &         0.90 \\
Canada         & 1,031,507 &         4.13 &   514,291 &         3.49 \\
Chile          &    75,879 &         0.30 &    49,256 &         0.33 \\
China          & 2,676,397 &        10.71 & 2,697,190 &        18.30 \\
Czechia        &   133,718 &         0.53 &    63,373 &         0.43 \\
Denmark        &   240,396 &         0.96 &   102,156 &         0.69 \\
Estonia        &    19,627 &         0.08 &     8,657 &         0.06 \\
Finland        &   194,681 &         0.78 &    82,959 &         0.56 \\
France         & 1,083,137 &         4.33 &   554,920 &         3.76 \\
Germany        & 1,506,443 &         6.03 &   811,283 &         5.50 \\
Greece         &   159,329 &         0.64 &    82,505 &         0.56 \\
Hungary        &    86,718 &         0.35 &    41,327 &         0.28 \\
Ireland        &   107,960 &         0.43 &    59,236 &         0.40 \\
Israel         &   230,267 &         0.92 &   109,401 &         0.74 \\
Italy          &   951,078 &         3.80 &   445,571 &         3.02 \\
Japan          & 1,400,348 &         5.60 &   974,167 &         6.61 \\
Lithuania      &    19,691 &         0.08 &    11,582 &         0.08 \\
Mexico         &   143,968 &         0.58 &   118,861 &         0.81 \\
Netherlands    &   589,909 &         2.36 &   259,720 &         1.76 \\
New Zealand    &   131,852 &         0.53 &    59,727 &         0.41 \\
Norway         &   171,731 &         0.69 &    69,874 &         0.47 \\
Poland         &   268,234 &         1.07 &   129,019 &         0.88 \\
Portugal       &   151,205 &         0.60 &    74,717 &         0.51 \\
Romania        &    52,317 &         0.21 &    28,475 &         0.19 \\
Russia         &   291,816 &         1.17 &   167,358 &         1.14 \\
Singapore      &   151,210 &         0.60 &    83,954 &         0.57 \\
Slovakia       &    37,755 &         0.15 &    20,473 &         0.14 \\
Slovenia       &    42,884 &         0.17 &    16,294 &         0.11 \\
South Africa   &   132,212 &         0.53 &    63,790 &         0.43 \\
South Korea    &   614,484 &         2.46 &   454,349 &         3.08 \\
Spain          &   741,808 &         2.97 &   403,012 &         2.73 \\
Sweden         &   397,464 &         1.59 &   163,304 &         1.11 \\
Switzerland    &   398,083 &         1.59 &   197,423 &         1.34 \\
Taiwan         &   354,338 &         1.42 &   255,617 &         1.73 \\
Turkey         &   241,956 &         0.97 &   158,887 &         1.08 \\
United Kingdom & 1,895,106 &         7.58 &   914,993 &         6.21 \\
United States  & 6,873,808 &        27.49 & 3,861,146 &        26.19 \\
\bottomrule
\end{tabular}
}
 \justify \small{\textit{Notes:} The table shows the number of authors and articles by country. Articles-columns contain duplicates if authored by authors from multiple countries. The country of an author is the country of the first-listed affiliation. Table excludes authors where the country is unknown.} 
\end{table}

\begin{figure}[H]
  \caption{Share of articles with usable affiliation information by field\label{fig:useable_articles}}
  \centering
  \includegraphics[width=0.57\textwidth]{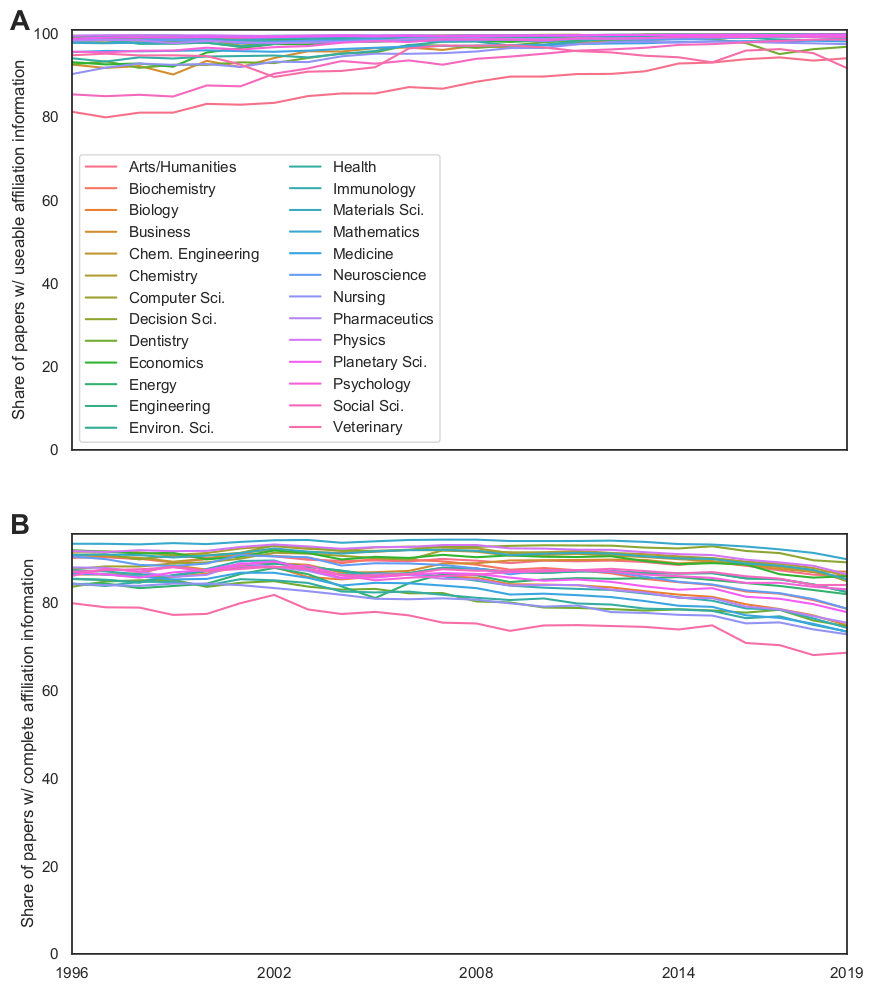}
  \justify \small{\textit{Notes:} The share of papers (research-type articles) with usable information. (A) Share of papers with existing author and affiliation information in the total number of papers published. (B) Share of papers thereof with where information for at least one organisation type for at least one author is available.}
\end{figure}

\begin{figure}[H]
  \caption{Share of authors with multiple affiliation by field over time accounting for countries' differing share in fields\label{fig:multiaff_fields-countryfield}}
  \centering
  \includegraphics[width=0.57\textwidth]{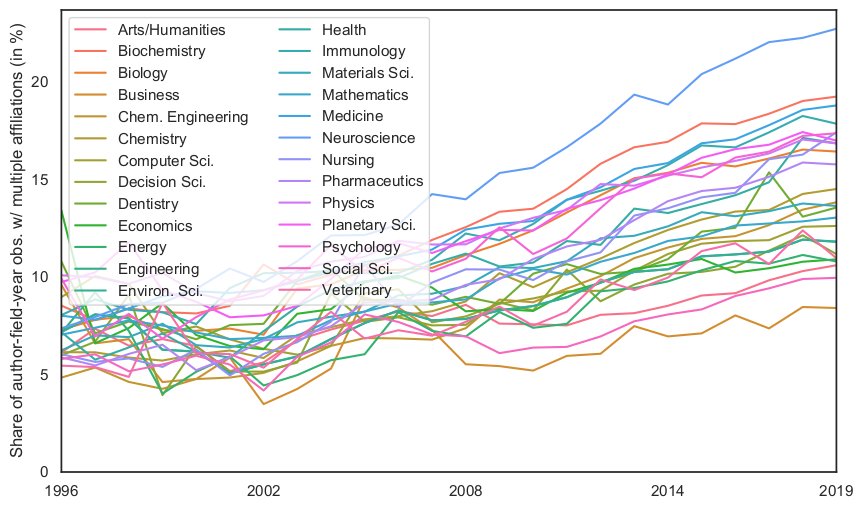}
  \justify \small{\textit{Notes:} The share of authors with multiple affiliations  $\overline{s}_{f,t}$.}
\end{figure}

\begin{table}[ht]
 \caption{Share of authors with multiple affiliations by field over time (in \%)
 \label{tab:multiaff_authors_share-field}}
 \resizebox{\columnwidth}{!}{%
 \centering
\begin{tabular}{lrrrrrrrrrrrrrrrrrrrrrrrr|r}
\toprule
{} &  1996 &  1997 &  1998 &  1999 &  2000 &  2001 &  2002 &  2003 &  2004 &  2005 &  2006 &  2007 &  2008 &  2009 &  2010 &  2011 &  2012 &  2013 &  2014 &  2015 &  2016 &  2017 &  2018 &  2019 &  Average \\
\midrule
Arts/Humanities   &   6.7 &   6.9 &   6.9 &   6.5 &   6.1 &   5.1 &   5.0 &   7.3 &   7.9 &   8.4 &   8.2 &   7.8 &   8.5 &   7.6 &   7.3 &   7.5 &   7.9 &   8.3 &   8.7 &   9.1 &   9.4 &   9.8 &  10.3 &  10.5 &      7.8 \\
Biochemistry      &  10.6 &  10.5 &  10.6 &  10.5 &  10.1 &  10.5 &  13.0 &  11.4 &  11.5 &  11.9 &  12.3 &  12.9 &  13.7 &  14.3 &  14.3 &  15.1 &  15.7 &  16.3 &  16.4 &  16.9 &  17.1 &  17.3 &  17.7 &  17.5 &     13.7 \\
Biology           &  10.9 &  10.6 &  10.9 &  10.2 &   9.4 &   8.4 &   7.9 &  11.1 &  11.4 &  11.9 &  11.7 &  11.9 &  12.6 &  13.0 &  13.3 &  14.2 &  14.9 &  15.4 &  15.4 &  15.8 &  16.0 &  16.2 &  16.6 &  16.5 &     12.8 \\
Business          &   6.1 &   7.0 &   6.5 &   4.9 &   5.1 &   5.7 &   3.6 &   4.3 &   6.1 &   7.9 &   7.9 &   6.3 &   6.2 &   5.1 &   4.3 &   5.0 &   5.3 &   5.9 &   6.2 &   6.6 &   6.8 &   7.2 &   7.7 &   7.9 &      6.1 \\
Chem. Engineering &   6.2 &   6.1 &   5.5 &   5.1 &   5.2 &   5.3 &   5.3 &   6.1 &   6.6 &   7.3 &   7.4 &   7.3 &   8.0 &   9.6 &   9.4 &   9.8 &  10.3 &  10.6 &  11.2 &  11.8 &  12.1 &  12.4 &  12.8 &  13.1 &      8.5 \\
Chemistry         &   7.4 &   7.5 &   6.9 &   6.3 &   6.4 &   6.4 &   6.2 &   7.3 &   7.7 &   8.3 &   8.5 &   8.7 &   9.1 &  10.7 &  10.1 &  10.3 &  10.9 &  11.5 &  12.1 &  12.8 &  13.1 &  13.5 &  13.9 &  13.9 &      9.6 \\
Computer Sci.     &   9.5 &  10.6 &  10.1 &   7.3 &   6.7 &   6.0 &   5.6 &   5.7 &   6.2 &   8.4 &   8.7 &   7.1 &   7.4 &   7.6 &   7.9 &   8.4 &   8.7 &   9.4 &   9.8 &  10.5 &  11.0 &  11.3 &  11.8 &  12.1 &      8.6 \\
Decision Sci.     &   6.8 &   7.7 &   8.6 &   4.8 &   5.3 &   6.8 &   4.4 &   5.4 &   7.2 &   8.5 &   7.7 &   7.3 &   7.0 &   7.2 &   6.4 &   7.2 &   7.4 &   8.1 &   8.6 &   8.8 &   9.3 &   9.9 &  10.6 &  10.5 &      7.6 \\
Dentistry         &  11.1 &   9.1 &   9.3 &   8.2 &   9.0 &   8.5 &   8.2 &   9.3 &   9.4 &   9.4 &   8.6 &   7.8 &   8.1 &   8.1 &   8.4 &   8.7 &   8.8 &   9.5 &   9.5 &  10.6 &   9.9 &  10.5 &  12.1 &  12.3 &      9.3 \\
Economics         &   6.8 &   5.6 &   6.1 &   5.8 &   6.5 &   7.0 &   6.1 &   6.8 &   8.8 &   9.1 &   9.2 &   9.0 &   9.0 &   8.1 &   7.9 &   8.7 &   9.3 &   9.4 &  10.2 &  10.2 &   9.7 &   9.8 &  10.3 &   9.9 &      8.3 \\
Energy            &   9.7 &  10.8 &   9.6 &   5.2 &   5.4 &   6.1 &   4.5 &   4.8 &   5.8 &   6.9 &   8.0 &   8.3 &   7.9 &   8.8 &   8.2 &   8.3 &   9.3 &   9.4 &   9.7 &  10.3 &  11.0 &  11.5 &  12.1 &  12.1 &      8.5 \\
Engineering       &   8.8 &   9.7 &   9.4 &   7.0 &   6.5 &   4.9 &   5.5 &   5.7 &   6.7 &   8.0 &   8.5 &   7.7 &   8.0 &   8.4 &   8.8 &   9.0 &   9.4 &   9.8 &  10.0 &  10.6 &  11.1 &  11.6 &  12.2 &  12.3 &      8.7 \\
Environ. Sci.     &   8.8 &   9.7 &   9.4 &   7.0 &   6.4 &   4.9 &   5.5 &   5.7 &   6.7 &   8.0 &   8.5 &   7.7 &   8.0 &   8.4 &   8.8 &   9.0 &   9.4 &   9.8 &  10.0 &  10.6 &  11.1 &  11.6 &  12.2 &  12.3 &      8.7 \\
Health            &   9.6 &   9.5 &   9.7 &   9.8 &   8.2 &   7.8 &   9.2 &   9.8 &  10.6 &  11.6 &  10.7 &  11.0 &  11.9 &  11.5 &  10.6 &  11.5 &  12.0 &  12.5 &  13.3 &  13.5 &  13.8 &  14.7 &  15.9 &  15.7 &     11.4 \\
Immunology        &  11.6 &  10.9 &  11.1 &  11.1 &   9.3 &  11.3 &  13.0 &  12.2 &  11.7 &  11.9 &  12.0 &  12.5 &  13.3 &  13.4 &  14.0 &  14.8 &  15.5 &  15.4 &  15.7 &  16.6 &  16.7 &  17.0 &  17.7 &  17.2 &     13.6 \\
Materials Sci.    &   7.5 &   8.0 &   7.7 &   8.0 &   7.0 &   6.5 &   7.0 &   7.6 &   8.2 &   8.6 &   9.3 &   9.4 &   9.7 &  10.7 &  10.6 &  11.0 &  11.2 &  11.6 &  12.0 &  12.6 &  12.7 &  13.3 &  13.5 &  13.4 &      9.9 \\
Mathematics       &   7.9 &   8.5 &   8.4 &   7.6 &   7.3 &   6.9 &   6.4 &   7.1 &   7.8 &   8.7 &   8.7 &   8.5 &   8.7 &   9.3 &   9.6 &   9.5 &  10.0 &  10.3 &  10.9 &  11.1 &  11.6 &  11.8 &  11.8 &  11.8 &      9.2 \\
Medicine          &  10.2 &  10.5 &  10.3 &  11.1 &  11.5 &  11.3 &  11.6 &  11.5 &  11.8 &  12.2 &  12.2 &  12.5 &  13.2 &  13.2 &  13.0 &  14.2 &  14.5 &  14.9 &  15.1 &  15.7 &  16.0 &  16.4 &  16.9 &  17.0 &     13.2 \\
Neuroscience      &  10.5 &  10.9 &  11.4 &  11.0 &  11.7 &  11.2 &  11.1 &  12.5 &  13.7 &  14.1 &  14.3 &  15.0 &  15.8 &  16.2 &  16.5 &  17.2 &  18.0 &  18.7 &  18.9 &  19.5 &  19.7 &  20.0 &  20.5 &  20.3 &     15.4 \\
Nursing           &   8.4 &   8.0 &   7.9 &   7.2 &   7.0 &   6.5 &   8.4 &   9.2 &   9.4 &  10.4 &  10.6 &  11.3 &  11.4 &  11.0 &  10.1 &  11.4 &  12.0 &  12.6 &  12.5 &  13.1 &  13.7 &  14.4 &  15.4 &  16.0 &     10.8 \\
Pharmaceutics     &   7.4 &   7.2 &   7.0 &   7.4 &   6.5 &   6.3 &   7.6 &   7.8 &   7.9 &   8.7 &   8.9 &   9.2 &   9.7 &   9.9 &  10.3 &  11.0 &  11.4 &  12.1 &  12.6 &  12.9 &  13.2 &  13.7 &  14.2 &  14.2 &      9.9 \\
Physics           &  11.3 &  11.2 &  10.5 &  10.7 &   9.4 &   9.2 &   9.2 &  10.2 &  10.8 &  11.4 &  11.8 &  11.1 &  11.6 &  12.3 &  12.7 &  13.1 &  13.7 &  13.8 &  14.2 &  14.7 &  15.1 &  15.6 &  15.8 &  15.7 &     12.3 \\
Planetary Sci.    &  13.5 &  14.5 &  13.9 &  12.0 &  10.5 &   9.7 &   9.2 &   9.2 &  11.0 &  13.1 &  12.9 &  12.8 &  13.1 &  13.8 &  14.0 &  14.6 &  15.3 &  15.5 &  16.4 &  17.0 &  17.6 &  18.2 &  18.9 &  18.9 &     14.0 \\
Psychology        &  11.4 &  10.4 &  10.1 &   9.0 &   9.8 &  10.2 &   9.6 &  11.3 &  14.0 &  14.0 &  13.3 &  11.6 &  11.7 &  10.5 &  10.7 &  11.7 &  12.0 &  13.6 &  14.1 &  14.7 &  14.9 &  15.1 &  15.6 &  15.6 &     12.3 \\
Social Sci.       &   6.8 &   6.3 &   6.1 &   5.4 &   5.5 &   5.4 &   4.1 &   5.5 &   7.4 &   8.4 &   8.0 &   7.0 &   6.8 &   5.8 &   5.5 &   5.8 &   6.2 &   6.8 &   7.5 &   7.8 &   8.4 &   8.8 &   9.3 &   9.3 &      6.8 \\
Veterinary        &   8.2 &   8.3 &   7.4 &   7.2 &   7.1 &   7.2 &   5.8 &   8.6 &   8.9 &   8.8 &   8.3 &   7.9 &   8.9 &   9.0 &   8.5 &   9.4 &  10.5 &  10.2 &  10.9 &  11.3 &  11.7 &  11.5 &  12.0 &  11.7 &      9.1 \\
\cmidrule{1-25}
All               &  10.9 &  10.9 &  10.8 &  10.6 &  10.4 &   9.8 &  10.4 &  10.7 &  11.2 &  11.9 &  12.0 &  11.9 &  12.4 &  12.6 &  12.4 &  13.3 &  13.7 &  14.1 &  14.4 &  14.9 &  15.3 &  15.6 &  16.0 &  15.9 &     12.6 \\
\bottomrule
\end{tabular}%
}
\justify \small{\textit{Notes:} The table shows the share $s_{f,t}$ of authors with multiple affiliations by field and year not accounting for countries' differing shares in fields. In each year we count unique authors based on their Scopus Author ID.}
\end{table}

\begin{table}[ht]
 \caption{Share of authors with multiple affiliations by journal quality group over time (in \%)
 \label{tab:multiaff_authors_share-quality}}
 \resizebox{\columnwidth}{!}{%
 \centering
\begin{tabular}{lrrrrrrrrrrrrrrrrrrrrrrrr|r}
\toprule
Journal quality group &  1996 &  1997 &  1998 &  1999 &  2000 &  2001 &  2002 &  2003 &  2004 &  2005 &  2006 &  2007 &  2008 &  2009 &  2010 &  2011 &  2012 &  2013 &  2014 &  2015 &  2016 &  2017 &  2018 &  2019 &  Average \\
\midrule
Top    &  10.5 &  10.7 &  10.8 &  11.4 &  11.0 &  10.8 &  11.8 &  11.3 &  11.8 &  12.4 &  12.3 &  12.4 &  13.1 &  13.6 &  13.4 &  14.4 &  14.7 &  15.1 &  15.7 &  16.4 &  16.7 &  17.0 &  17.6 &  17.4 &     13.4 \\
Second &   9.7 &   9.3 &   9.1 &   8.3 &   8.5 &   7.7 &   7.6 &   8.7 &   9.2 &  10.1 &  10.2 &  10.1 &  10.7 &  10.6 &  10.8 &  11.6 &  12.2 &  12.8 &  12.9 &  13.3 &  13.6 &  13.9 &  14.4 &  14.3 &     10.8 \\
Third  &   7.6 &   7.9 &   7.5 &   7.0 &   6.3 &   5.8 &   5.9 &   7.3 &   7.9 &   8.3 &   8.6 &   8.5 &   8.9 &   8.9 &   9.1 &   9.8 &  10.0 &  10.3 &  10.8 &  11.2 &  11.7 &  11.9 &  12.3 &  12.5 &      9.0 \\
Fourth &   7.2 &   7.4 &   6.9 &   6.2 &   6.1 &   5.4 &   5.5 &   6.9 &   7.4 &   7.9 &   8.1 &   8.1 &   8.3 &   8.2 &   8.2 &   8.5 &   8.9 &   9.0 &   9.2 &   9.7 &  10.0 &  10.5 &  11.0 &  11.3 &      8.2 \\
\bottomrule
\end{tabular}%
}
\justify \small{\textit{Notes:} The table shows the share of authors with multiple affiliations by journal quality group and year. In each year we count unique authors based on their Scopus Author ID.}
\end{table}

\begin{table}[ht]
 \caption{Share of authors with multiple affiliations by country over time (in \%)
 \label{tab:multiaff_authors_share-country}}
 \resizebox{\columnwidth}{!}{%
 \centering
\begin{tabular}{lrrrrrrrrrrrrrrrrrrrrrrrr|r}
\toprule
{} &  1996 &  1997 &  1998 &  1999 &  2000 &  2001 &  2002 &  2003 &  2004 &  2005 &  2006 &  2007 &  2008 &  2009 &  2010 &  2011 &  2012 &  2013 &  2014 &  2015 &  2016 &  2017 &  2018 &  2019 &  Average \\
\midrule
Argentina      &  10.6 &   8.4 &   9.7 &   9.3 &   9.2 &   9.8 &  11.1 &  11.9 &  13.6 &  14.3 &  14.8 &  15.3 &  16.5 &  17.6 &  18.3 &  18.9 &  19.3 &  20.1 &  21.1 &  20.9 &  21.7 &  22.6 &  24.8 &  25.2 &     16.0 \\
Australia      &  11.7 &  12.3 &  12.2 &  11.6 &  11.1 &  10.9 &  11.6 &  14.1 &  15.3 &  16.3 &  16.9 &  17.0 &  17.9 &  17.6 &  17.8 &  19.5 &  20.5 &  21.1 &  21.5 &  22.4 &  23.1 &  24.5 &  25.5 &  25.0 &     17.4 \\
Austria        &   9.5 &   9.0 &   9.9 &   9.5 &   9.5 &   9.5 &   9.6 &  10.2 &  10.9 &  11.2 &  10.2 &  10.3 &  10.8 &  11.4 &  11.0 &  11.7 &  12.7 &  12.6 &  13.6 &  14.2 &  14.2 &  14.2 &  14.9 &  15.3 &     11.5 \\
Belgium        &  10.2 &  10.2 &  10.1 &   9.9 &   9.3 &   9.5 &   9.9 &  10.1 &  11.3 &  11.6 &  11.7 &  12.3 &  13.7 &  14.7 &  14.5 &  15.3 &  15.9 &  17.9 &  18.3 &  18.9 &  19.4 &  20.2 &  21.3 &  21.8 &     14.1 \\
Canada         &  14.6 &  15.0 &  14.2 &  13.9 &  13.3 &  12.2 &  12.9 &  14.8 &  14.9 &  16.4 &  15.8 &  15.7 &  15.9 &  16.0 &  15.4 &  16.9 &  17.4 &  17.4 &  18.1 &  19.0 &  19.3 &  20.0 &  20.4 &  20.0 &     16.2 \\
Chile          &   7.3 &   5.7 &   5.8 &   6.7 &   5.8 &   7.0 &   6.7 &   7.8 &   8.9 &  10.2 &  10.9 &  11.0 &  11.7 &  11.9 &  11.8 &  11.8 &  13.2 &  13.6 &  14.4 &  15.9 &  15.5 &  17.7 &  17.6 &  18.3 &     11.1 \\
China          &   5.5 &   5.3 &   5.5 &   5.6 &   6.1 &   5.8 &   6.4 &   7.8 &   9.0 &  10.3 &  11.5 &  12.0 &  12.2 &  12.6 &  12.5 &  12.8 &  12.8 &  13.0 &  13.2 &  13.8 &  14.6 &  15.3 &  15.9 &  16.0 &     10.6 \\
Czechia        &   6.8 &   6.7 &   6.9 &   7.5 &   8.0 &   7.6 &   7.8 &   9.5 &   9.8 &  10.4 &  11.3 &  11.8 &  12.4 &  12.6 &  12.1 &  12.3 &  13.8 &  14.1 &  14.1 &  15.5 &  15.7 &  16.4 &  17.3 &  17.6 &     11.6 \\
Denmark        &  11.5 &  11.5 &  11.1 &  11.6 &  11.5 &  11.2 &  11.0 &  11.7 &  12.2 &  12.5 &  12.1 &  12.5 &  13.1 &  13.2 &  13.7 &  14.1 &  14.2 &  15.5 &  16.0 &  17.0 &  17.4 &  17.7 &  19.0 &  19.8 &     13.8 \\
Estonia        &   5.9 &   5.7 &   9.8 &   7.8 &   8.4 &   6.3 &   7.9 &   9.6 &  10.3 &   9.5 &  11.8 &  10.0 &  11.3 &  10.9 &  12.4 &  15.6 &  13.9 &  15.0 &  15.8 &  17.4 &  15.1 &  15.2 &  14.8 &  14.4 &     11.4 \\
Finland        &  11.3 &  11.2 &  12.5 &  12.6 &  12.3 &  11.7 &  12.0 &  14.1 &  14.4 &  14.5 &  14.9 &  14.8 &  15.1 &  15.8 &  15.1 &  16.0 &  16.9 &  16.7 &  17.7 &  17.8 &  18.7 &  19.2 &  20.0 &  20.7 &     15.3 \\
France         &   9.4 &   9.8 &  10.3 &  10.7 &  10.2 &  10.0 &  10.8 &  11.1 &  11.9 &  13.1 &  15.5 &  18.4 &  21.2 &  23.6 &  24.6 &  26.8 &  28.9 &  29.7 &  30.9 &  31.4 &  29.8 &  27.3 &  26.0 &  24.3 &     19.4 \\
Germany        &   9.6 &   9.6 &   9.5 &   9.8 &   9.1 &   9.1 &   9.8 &  10.0 &  10.5 &  11.0 &  11.1 &  11.0 &  11.5 &  12.1 &  12.2 &  12.8 &  13.3 &  14.1 &  14.6 &  15.2 &  15.8 &  16.2 &  17.4 &  17.4 &     12.2 \\
Greece         &   7.6 &   7.4 &   7.4 &   7.0 &   7.6 &   6.9 &   7.1 &   7.6 &   6.7 &   8.2 &   8.0 &   7.7 &   8.2 &   8.3 &   7.9 &   8.3 &   8.9 &   9.5 &   9.4 &   9.8 &  10.7 &  11.0 &  10.7 &  11.0 &      8.5 \\
Hungary        &   5.4 &   6.8 &   6.9 &   7.3 &   7.6 &   7.4 &   8.4 &   7.6 &   8.0 &   8.4 &   8.4 &   8.7 &   8.9 &   8.8 &   9.0 &   9.4 &   9.9 &  11.5 &  13.1 &  12.8 &  14.3 &  14.4 &  15.9 &  16.0 &      9.8 \\
Ireland        &   8.1 &   7.5 &   8.1 &   8.1 &   7.1 &   7.2 &   8.6 &   8.5 &   9.5 &  10.1 &  10.5 &   9.5 &  10.6 &  12.1 &  10.9 &  11.8 &  13.3 &  13.3 &  14.1 &  15.0 &  14.4 &  14.6 &  16.3 &  15.9 &     11.0 \\
Israel         &  12.7 &  12.7 &  12.8 &  12.5 &  13.8 &  12.2 &  13.3 &  14.4 &  15.4 &  16.4 &  15.4 &  15.2 &  14.6 &  14.3 &  14.2 &  14.9 &  15.6 &  16.5 &  16.5 &  16.9 &  17.9 &  19.0 &  19.2 &  19.1 &     15.2 \\
Italy          &   8.3 &   8.4 &   8.4 &   9.1 &   9.1 &   9.0 &  10.0 &   8.2 &   9.0 &   9.6 &   9.9 &   9.4 &  10.5 &  10.2 &  10.1 &  10.6 &  10.8 &  11.4 &  11.5 &  12.4 &  12.6 &  12.8 &  13.5 &  14.1 &     10.4 \\
Japan          &   7.1 &   7.3 &   7.3 &   7.9 &   8.2 &   7.9 &   8.8 &   8.5 &   9.0 &   9.6 &   9.4 &   9.5 &   9.8 &   9.7 &  10.1 &  10.3 &  10.2 &  10.4 &  10.3 &  11.0 &  11.2 &  11.3 &  11.8 &  11.6 &      9.5 \\
Lithuania      &   8.2 &   5.9 &   6.3 &   9.3 &   4.9 &   5.6 &   5.2 &   5.4 &   4.9 &   5.9 &   6.7 &   4.6 &   5.6 &   5.4 &   5.5 &   6.2 &   6.8 &   7.6 &   7.3 &   6.9 &   7.3 &   7.4 &   7.9 &   7.2 &      6.4 \\
Mexico         &   8.4 &   7.7 &   7.7 &   8.2 &   8.0 &   8.6 &   8.4 &   8.3 &   8.6 &   8.1 &   8.4 &   7.8 &   7.5 &   7.1 &   6.4 &   7.1 &   6.8 &   6.7 &   7.0 &   7.4 &   7.0 &   7.3 &   7.7 &   7.7 &      7.7 \\
Netherlands    &  10.7 &  11.4 &  10.8 &  11.2 &  11.7 &  11.6 &  12.1 &  12.1 &  12.7 &  13.8 &  14.1 &  13.7 &  15.1 &  15.3 &  15.5 &  16.9 &  17.2 &  17.6 &  18.0 &  18.4 &  18.4 &  19.2 &  19.6 &  19.7 &     14.9 \\
New Zealand    &   8.3 &   8.8 &   9.5 &   7.3 &   7.7 &   7.2 &   7.6 &   8.7 &   9.5 &   9.9 &  10.1 &  10.8 &   9.9 &  10.6 &   9.8 &  11.3 &  11.7 &  12.0 &  13.0 &  13.4 &  13.5 &  13.4 &  14.4 &  14.5 &     10.5 \\
Norway         &  10.9 &  11.3 &  11.2 &  10.9 &  10.1 &  10.0 &  10.8 &  11.0 &  12.7 &  14.6 &  16.4 &  19.5 &  19.9 &  21.5 &  21.3 &  22.2 &  23.5 &  23.6 &  24.5 &  24.7 &  25.5 &  25.1 &  25.5 &  24.7 &     18.0 \\
Poland         &   4.6 &   5.1 &   5.1 &   5.6 &   5.2 &   5.3 &   5.2 &   5.7 &   6.2 &   6.3 &   6.3 &   6.2 &   6.1 &   6.4 &   6.1 &   6.5 &   6.1 &   6.6 &   6.3 &   6.8 &   6.8 &   6.9 &   7.2 &   6.9 &      6.1 \\
Portugal       &  10.7 &   9.3 &   9.9 &  11.1 &  11.9 &  11.3 &  12.2 &  12.7 &  14.1 &  14.8 &  15.1 &  15.5 &  16.7 &  17.7 &  18.0 &  19.6 &  20.5 &  22.4 &  23.0 &  24.7 &  25.1 &  25.8 &  27.4 &  27.4 &     17.4 \\
Romania        &   3.7 &   4.1 &   4.6 &   6.1 &   6.3 &   8.2 &   5.8 &   6.0 &   6.2 &   6.0 &   6.8 &   6.5 &   5.9 &   6.7 &   7.1 &   7.9 &   9.1 &  10.2 &  11.3 &  12.5 &  12.8 &  13.2 &  14.0 &  14.2 &      8.1 \\
Russia         &   4.5 &   4.4 &   4.5 &   5.1 &   5.2 &   4.6 &   5.6 &   6.1 &   6.6 &   7.1 &   7.7 &   6.4 &   8.0 &   8.7 &  10.7 &  11.5 &  13.6 &  15.7 &  17.7 &  18.8 &  19.4 &  19.5 &  19.9 &  20.2 &     10.5 \\
Singapore      &   6.7 &   7.3 &   9.0 &   6.5 &   6.4 &   6.9 &   7.2 &   8.7 &   9.5 &  11.0 &  12.2 &  11.7 &  13.0 &  12.9 &  12.3 &  13.5 &  15.2 &  16.5 &  17.1 &  18.9 &  19.4 &  18.8 &  19.7 &  19.7 &     12.5 \\
Slovakia       &   4.9 &   4.3 &   5.1 &   5.6 &   5.8 &   5.4 &   5.2 &   6.2 &   5.2 &   6.5 &   7.4 &   6.0 &   7.4 &   7.2 &   7.9 &   7.8 &   6.7 &   8.3 &   7.7 &   8.8 &   8.2 &   8.2 &   8.3 &   9.0 &      6.8 \\
Slovenia       &   7.0 &   6.9 &   7.8 &   7.9 &   7.8 &   5.4 &   6.5 &   7.5 &   6.8 &   7.5 &   7.6 &   7.7 &   8.2 &   7.3 &   8.3 &  11.2 &  13.1 &  13.3 &  11.7 &  12.0 &  12.8 &  12.4 &  13.7 &  14.1 &      9.4 \\
South Africa   &  10.3 &   9.6 &   9.2 &   8.7 &   7.6 &   7.4 &   7.6 &   9.6 &  10.4 &  11.6 &  10.8 &  12.1 &  12.2 &  12.7 &  12.7 &  14.0 &  13.8 &  15.5 &  15.9 &  16.8 &  17.7 &  18.4 &  19.0 &  18.8 &     12.6 \\
South Korea    &   5.6 &   5.8 &   5.6 &   5.7 &   5.8 &   5.6 &   5.5 &   6.0 &   6.3 &   6.8 &   7.4 &   6.6 &   6.8 &   6.9 &   6.9 &   7.5 &   7.7 &   8.3 &   8.8 &   9.2 &   9.8 &  10.0 &  10.3 &  10.1 &      7.3 \\
Spain          &   6.2 &   7.1 &   6.5 &   6.5 &   6.8 &   5.9 &   6.5 &   6.4 &   7.0 &   7.3 &   7.9 &   8.2 &  10.1 &  11.0 &  11.4 &  12.4 &  13.3 &  14.2 &  15.0 &  15.7 &  16.3 &  17.1 &  18.3 &  18.8 &     10.7 \\
Sweden         &  11.6 &  12.0 &  11.7 &  12.4 &  11.6 &  11.5 &  12.7 &  13.1 &  13.7 &  14.3 &  14.8 &  14.7 &  15.4 &  15.8 &  15.3 &  16.2 &  16.6 &  17.5 &  17.8 &  18.2 &  19.4 &  20.3 &  20.5 &  21.5 &     15.4 \\
Switzerland    &  11.6 &  11.2 &  11.6 &  11.8 &  11.1 &  10.2 &  10.8 &  11.3 &  11.9 &  12.5 &  12.6 &  12.8 &  13.1 &  13.4 &  13.7 &  14.6 &  16.6 &  17.1 &  16.9 &  17.1 &  17.4 &  18.3 &  18.6 &  18.9 &     14.0 \\
Taiwan         &   7.4 &   8.2 &   8.2 &   8.4 &   8.4 &   7.5 &   8.6 &   9.8 &  10.0 &  11.0 &  12.4 &  12.5 &  13.0 &  13.5 &  13.6 &  14.4 &  15.2 &  15.4 &  16.1 &  17.2 &  17.7 &  18.8 &  19.3 &  19.7 &     12.8 \\
Turkey         &   5.2 &   4.9 &   5.1 &   5.2 &   5.7 &   5.0 &   4.6 &   5.0 &   4.8 &   4.7 &   3.8 &   3.6 &   3.5 &   3.3 &   2.5 &   2.7 &   3.0 &   2.8 &   3.2 &   3.3 &   3.3 &   3.6 &   4.2 &   4.4 &      4.1 \\
United Kingdom &  11.0 &  10.5 &  10.2 &   9.7 &   9.0 &   8.8 &   8.6 &   9.9 &  10.4 &  11.1 &  10.3 &   9.7 &  10.3 &  10.4 &  10.0 &  11.2 &  11.9 &  12.6 &  13.3 &  13.7 &  14.3 &  14.8 &  15.5 &  15.4 &     11.3 \\
United States  &  13.8 &  14.0 &  13.8 &  13.2 &  12.8 &  11.7 &  12.3 &  12.9 &  13.5 &  14.2 &  13.8 &  13.1 &  13.2 &  12.8 &  12.2 &  13.2 &  13.4 &  13.6 &  13.7 &  14.1 &  14.3 &  14.2 &  14.6 &  14.1 &     13.4 \\
\bottomrule
\end{tabular}%
}
\justify \small{\textit{Notes:} The table shows the share of authors $s_{c,t}$ with multiple affiliations by country and year not accounting for countries' differing shares in fields. In each year we count unique authors based on their Scopus Author ID.}
\end{table}

\newpage
\begin{table}[ht!]
  \caption{Organisation types as share of all author-article observations by year (in \%)\label{tab:afftype_share-all_comparison}}
  \resizebox{\columnwidth}{!}{%
  \centering
  \begin{tabular}{lrrrrrrrrrrrrrrrrrrrrrrrr}
\toprule
{} &  1996 &  1997 &  1998 &  1999 &  2000 &  2001 &  2002 &  2003 &  2004 &  2005 &  2006 &  2007 &  2008 &  2009 &  2010 &  2011 &  2012 &  2013 &  2014 &  2015 &  2016 &  2017 &  2018 &  2019 \\
\midrule
Univ.                 &  55.3 &  54.7 &  55.2 &  55.1 &  55.7 &  56.4 &  56.7 &  57.1 &  56.9 &  56.9 &  57.3 &  57.7 &  57.9 &  57.7 &  58.2 &  58.2 &  58.2 &  58.3 &  58.6 &  58.6 &  58.7 &  59.0 &  59.4 &  60.0 \\
Res. Inst.            &  14.3 &  14.7 &  14.8 &  15.0 &  14.9 &  15.2 &  15.1 &  14.7 &  14.5 &  14.4 &  14.3 &  14.2 &  13.9 &  13.5 &  13.3 &  13.2 &  12.8 &  12.5 &  12.1 &  11.8 &  11.4 &  10.9 &  10.3 &   9.6 \\
Unknown               &   5.8 &   5.9 &   6.1 &   6.6 &   6.3 &   5.7 &   5.4 &   5.4 &   5.6 &   5.2 &   5.1 &   5.0 &   5.1 &   5.6 &   5.5 &   5.2 &   5.2 &   5.4 &   5.5 &   5.4 &   5.7 &   5.7 &   5.8 &   6.3 \\
Hospital              &   8.0 &   7.9 &   7.8 &   7.5 &   7.8 &   7.7 &   7.6 &   7.3 &   7.2 &   7.2 &   7.0 &   6.9 &   6.7 &   6.7 &   6.7 &   6.6 &   6.7 &   6.7 &   6.6 &   6.6 &   6.4 &   6.4 &   6.2 &   6.1 \\
Company               &   5.3 &   5.2 &   4.9 &   4.9 &   4.6 &   4.7 &   4.5 &   4.5 &   4.3 &   4.2 &   3.9 &   3.9 &   3.6 &   3.4 &   3.3 &   3.1 &   2.9 &   2.7 &   2.5 &   2.5 &   2.4 &   2.3 &   2.1 &   1.9 \\
Gov.                  &   1.3 &   1.3 &   1.3 &   1.3 &   1.3 &   1.3 &   1.3 &   1.2 &   1.2 &   1.2 &   1.3 &   1.3 &   1.3 &   1.3 &   1.2 &   1.2 &   1.1 &   1.1 &   1.1 &   1.0 &   1.0 &   1.0 &   1.0 &   0.9 \\
Non-Gov.              &   0.5 &   0.5 &   0.5 &   0.5 &   0.5 &   0.5 &   0.5 &   0.5 &   0.4 &   0.4 &   0.4 &   0.5 &   0.5 &   0.4 &   0.4 &   0.4 &   0.4 &   0.4 &   0.4 &   0.4 &   0.4 &   0.4 &   0.3 &   0.3 \\
Univ.-Res. Inst.      &   1.9 &   1.9 &   1.9 &   2.0 &   1.9 &   1.9 &   2.1 &   2.3 &   2.4 &   2.6 &   2.8 &   3.0 &   3.1 &   3.4 &   3.4 &   3.6 &   3.7 &   3.8 &   3.9 &   4.0 &   4.0 &   4.0 &   4.0 &   3.8 \\
Univ.-Univ.           &   2.4 &   2.3 &   2.3 &   2.2 &   2.1 &   2.0 &   2.1 &   2.3 &   2.4 &   2.6 &   2.5 &   2.5 &   2.6 &   2.7 &   2.7 &   2.8 &   3.0 &   3.1 &   3.1 &   3.3 &   3.3 &   3.4 &   3.6 &   3.5 \\
Other                 &   2.5 &   2.6 &   2.5 &   2.2 &   2.2 &   2.2 &   2.3 &   2.0 &   2.0 &   2.2 &   2.2 &   2.1 &   2.2 &   2.2 &   2.1 &   2.2 &   2.3 &   2.4 &   2.4 &   2.5 &   2.6 &   2.6 &   2.7 &   2.7 \\
Univ.-Unknown         &   1.1 &   1.2 &   1.1 &   1.0 &   0.9 &   0.8 &   0.8 &   0.9 &   1.1 &   1.2 &   1.2 &   1.0 &   1.1 &   1.2 &   1.2 &   1.2 &   1.3 &   1.5 &   1.6 &   1.7 &   1.9 &   2.1 &   2.3 &   2.5 \\
Univ.-Hospital        &   1.3 &   1.2 &   1.3 &   1.2 &   1.2 &   1.2 &   1.3 &   1.3 &   1.3 &   1.4 &   1.4 &   1.5 &   1.5 &   1.5 &   1.5 &   1.7 &   1.7 &   1.8 &   1.8 &   1.8 &   1.8 &   1.9 &   1.9 &   1.9 \\
Res. Inst.-Res. Inst. &   0.4 &   0.4 &   0.4 &   0.4 &   0.4 &   0.4 &   0.4 &   0.4 &   0.5 &   0.5 &   0.5 &   0.5 &   0.5 &   0.5 &   0.5 &   0.5 &   0.5 &   0.5 &   0.5 &   0.5 &   0.5 &   0.4 &   0.4 &   0.4 \\
\bottomrule
\end{tabular}%
}
  \justify \small{\textit{Notes:} The table shows the share of the respective affiliation (combination) type on all author-article observations. Organisation types as defined in Scopus with few manual aggregations.}
\end{table}

\begin{table}[ht!]
  \caption{Organisation type combinations by field (in \%)\label{tab:combs_share-field}}
  \resizebox{\columnwidth}{!}{%
  \centering
  {\scriptsize \begin{tabular}{lrrrrrrrrrrrr}
\toprule
{} &  Univ.- &  Univ.- &  Univ.- &  Univ.- &  Hospital- &  Hospital- &  Res. Inst.- &  Res. Inst.- &  Res. Inst.- &  Univ.- &  Univ.- &  Other \\
{} &  Univ. &  Res. Inst. &  Unknown &  Hospital &  Hospital &  Unknown &  Hospital &  Res. Inst. &  Unknown &  Company &  Gov. &  \\
\midrule
Arts/Humanities   &         33.3 &              22.9 &           15.0 &            10.6 &                 <3 &                <3 &                   <3 &                     <3 &                  <3 &             <3 &          <3 &    7.4 \\
Biochemistry      &         22.2 &              26.8 &            9.6 &            17.5 &                 <3 &                <3 &                   <3 &                    3.9 &                  <3 &             <3 &          <3 &    8.3 \\
Biology           &         25.7 &              32.2 &           11.9 &             6.2 &                 <3 &                <3 &                   <3 &                    5.0 &                 3.4 &             <3 &         3.5 &    8.3 \\
Business          &         49.1 &              17.0 &           19.8 &              <3 &                 <3 &                <3 &                   <3 &                     <3 &                  <3 &             <3 &          <3 &    5.5 \\
Chem. Engineering &         28.0 &              36.3 &           13.6 &             4.1 &                 <3 &                <3 &                   <3 &                    3.8 &                  <3 &            3.2 &          <3 &    5.4 \\
Chemistry         &         28.4 &              37.9 &           11.3 &             3.4 &                 <3 &                <3 &                   <3 &                    4.4 &                  <3 &            3.3 &          <3 &    6.1 \\
Computer Sci.     &         33.2 &              27.5 &           15.2 &             4.7 &                 <3 &                <3 &                   <3 &                    3.0 &                  <3 &            4.2 &          <3 &    6.4 \\
Decision Sci.     &         46.1 &              25.4 &           13.9 &              <3 &                 <3 &                <3 &                   <3 &                     <3 &                  <3 &             <3 &          <3 &    3.9 \\
Dentistry         &         26.7 &               9.6 &           21.7 &            24.4 &                 <3 &               3.3 &                   <3 &                     <3 &                  <3 &             <3 &          <3 &    6.4 \\
Economics         &         39.7 &              28.6 &           14.7 &              <3 &                 <3 &                <3 &                   <3 &                     <3 &                  <3 &             <3 &         3.7 &    7.5 \\
Energy            &         24.0 &              34.7 &           15.9 &              <3 &                 <3 &                <3 &                   <3 &                    5.1 &                 3.7 &            4.0 &          <3 &    9.6 \\
Engineering       &         27.3 &              29.5 &           17.5 &             3.4 &                 <3 &                <3 &                   <3 &                    3.6 &                 3.0 &            4.6 &          <3 &    7.5 \\
Environ. Sci.     &         27.3 &              29.5 &           17.4 &             3.4 &                 <3 &                <3 &                   <3 &                    3.7 &                 3.0 &            4.6 &          <3 &    7.5 \\
Health            &         20.6 &              13.5 &           14.5 &            27.8 &                 <3 &               3.5 &                  3.1 &                     <3 &                  <3 &             <3 &          <3 &    9.1 \\
Immunology        &         19.3 &              26.2 &            9.8 &            16.4 &                 <3 &                <3 &                  3.4 &                    4.2 &                  <3 &             <3 &          <3 &   10.8 \\
Materials Sci.    &         27.4 &              39.1 &           11.9 &              <3 &                 <3 &                <3 &                   <3 &                    5.2 &                  <3 &            3.6 &          <3 &    5.5 \\
Mathematics       &         40.0 &              32.2 &           10.9 &              <3 &                 <3 &                <3 &                   <3 &                    4.0 &                  <3 &             <3 &          <3 &    4.2 \\
Medicine          &         18.0 &              17.3 &           12.3 &            26.0 &                3.1 &               3.6 &                  3.4 &                     <3 &                  <3 &             <3 &          <3 &    9.4 \\
Neuroscience      &         22.4 &              22.6 &           10.0 &            22.8 &                 <3 &                <3 &                  3.6 &                     <3 &                  <3 &             <3 &          <3 &    8.5 \\
Nursing           &         18.8 &              13.9 &           15.6 &            28.1 &                 <3 &               3.2 &                  3.2 &                     <3 &                  <3 &             <3 &          <3 &    9.6 \\
Pharmaceutics     &         21.2 &              22.0 &           12.9 &            18.2 &                 <3 &                <3 &                   <3 &                     <3 &                  <3 &            3.0 &          <3 &   11.2 \\
Physics           &         24.5 &              43.9 &            8.3 &              <3 &                 <3 &                <3 &                   <3 &                    7.9 &                  <3 &             <3 &          <3 &    5.5 \\
Planetary Sci.    &         21.5 &              40.1 &           11.5 &              <3 &                 <3 &                <3 &                   <3 &                    8.5 &                 4.4 &             <3 &         3.7 &    8.0 \\
Psychology        &         28.8 &              15.6 &           15.2 &            22.4 &                 <3 &                <3 &                   <3 &                     <3 &                  <3 &             <3 &          <3 &    8.5 \\
Social Sci.       &         35.6 &              18.2 &           20.5 &             8.5 &                 <3 &                <3 &                   <3 &                     <3 &                  <3 &             <3 &          <3 &    7.9 \\
Veterinary        &         25.2 &              21.1 &           20.2 &             7.0 &                 <3 &                <3 &                   <3 &                     <3 &                 4.1 &             <3 &         3.6 &   10.8 \\
\bottomrule
\end{tabular}%
}
}
  \justify \small{\textit{Notes:} The table shows the share of the respective multiple affiliation combination on all author-article observations per field. Organisation type as defined in Scopus with few manual aggregations.}
\end{table}

\begin{table}[ht!]
  \caption{Share of authors with international co-affiliation by field over time (in \%) \label{tab:foreignaff_authors_share-countryfield}}
  \resizebox{\columnwidth}{!}{%
  \centering
  {\scriptsize \begin{tabular}{lrrrrrrrrrrrrrrrrrrrrrrrr|r}
\toprule
{} &  1996 &  1997 &  1998 &  1999 &  2000 &  2001 &  2002 &  2003 &  2004 &  2005 &  2006 &  2007 &  2008 &  2009 &  2010 &  2011 &  2012 &  2013 &  2014 &  2015 &  2016 &  2017 &  2018 &  2019 &  Average \\
\midrule
Arts/Humanities   &  49.9 &  31.3 &  38.5 &  42.3 &  39.6 &  37.4 &  38.7 &  37.0 &  37.3 &  38.7 &  35.0 &  31.9 &  34.4 &  31.5 &  32.3 &  35.5 &  34.0 &  36.4 &  34.7 &  34.9 &  33.2 &  36.4 &  35.8 &  36.8 &     36.4 \\
Biochemistry      &  41.4 &  41.1 &  40.6 &  35.6 &  38.3 &  40.4 &  43.1 &  35.0 &  35.1 &  28.9 &  27.8 &  28.9 &  27.0 &  25.6 &  26.7 &  27.0 &  25.5 &  25.6 &  24.4 &  24.1 &  24.2 &  23.8 &  23.6 &  22.6 &     30.7 \\
Biology           &  45.5 &  42.2 &  45.9 &  45.0 &  45.8 &  44.1 &  43.4 &  41.4 &  42.0 &  38.3 &  39.8 &  38.0 &  37.0 &  33.7 &  35.8 &  32.6 &  30.6 &  30.0 &  30.5 &  30.3 &  31.5 &  30.9 &  31.2 &  31.6 &     37.4 \\
Business          &  62.3 &  54.7 &  54.1 &  57.1 &  57.1 &  54.8 &  51.7 &  53.3 &  45.1 &  42.8 &  44.4 &  54.1 &  50.4 &  41.3 &  53.9 &  50.2 &  56.1 &  48.7 &  52.3 &  49.3 &  46.3 &  51.5 &  50.4 &  54.8 &     51.5 \\
Chem. Engineering &  44.2 &  44.5 &  43.3 &  42.4 &  43.5 &  38.2 &  40.4 &  43.7 &  39.7 &  36.2 &  34.6 &  32.2 &  32.3 &  34.5 &  31.5 &  30.5 &  31.1 &  31.2 &  29.7 &  30.5 &  30.9 &  30.7 &  31.4 &  29.2 &     35.7 \\
Chemistry         &  53.6 &  57.3 &  49.6 &  46.8 &  47.5 &  47.5 &  45.4 &  45.2 &  43.1 &  40.5 &  39.8 &  36.4 &  37.4 &  38.1 &  35.4 &  34.2 &  33.7 &  33.2 &  31.8 &  32.5 &  33.0 &  32.8 &  31.3 &  31.2 &     39.9 \\
Computer Sci.     &  58.6 &  60.2 &  59.4 &  48.8 &  53.3 &  46.8 &  42.3 &  47.8 &  44.5 &  38.5 &  35.6 &  37.9 &  37.7 &  36.9 &  36.1 &  34.6 &  36.0 &  35.9 &  36.0 &  34.3 &  36.1 &  35.0 &  34.3 &  34.0 &     41.7 \\
Decision Sci.     &  59.7 &  49.8 &  51.9 &  59.8 &  46.0 &  55.1 &  45.6 &  49.3 &  47.4 &  38.8 &  49.4 &  48.5 &  50.2 &  42.6 &  47.7 &  47.4 &  50.9 &  45.0 &  47.4 &  44.2 &  47.4 &  44.1 &  41.4 &  49.3 &     48.3 \\
Dentistry         &  29.6 &  26.9 &  24.9 &  24.0 &  26.6 &  29.8 &  28.1 &  28.1 &  22.0 &  19.3 &  19.9 &  26.3 &  29.6 &  22.2 &  33.4 &  27.5 &  27.0 &  29.7 &  31.3 &  24.7 &  28.0 &  25.2 &  31.2 &  23.1 &     26.6 \\
Economics         &  65.7 &  64.9 &  64.6 &  61.0 &  59.4 &  60.8 &  56.8 &  57.3 &  54.3 &  54.8 &  52.3 &  58.2 &  57.1 &  51.5 &  50.7 &  50.9 &  52.0 &  48.7 &  49.6 &  49.0 &  50.5 &  49.7 &  51.7 &  52.8 &     55.2 \\
Energy            &  61.4 &  55.3 &  57.1 &  42.1 &  54.7 &  53.4 &  42.6 &  44.3 &  37.8 &  47.9 &  43.7 &  47.4 &  46.0 &  44.3 &  35.2 &  39.3 &  35.7 &  33.6 &  32.8 &  32.9 &  35.5 &  36.4 &  38.9 &  37.8 &     43.2 \\
Engineering       &  59.2 &  54.7 &  50.8 &  48.3 &  52.8 &  42.1 &  41.7 &  41.4 &  39.0 &  38.1 &  36.2 &  33.0 &  33.0 &  30.5 &  32.9 &  30.3 &  32.6 &  31.3 &  33.2 &  30.6 &  32.8 &  33.2 &  33.7 &  33.3 &     38.5 \\
Environ. Sci.     &  59.1 &  54.6 &  50.9 &  48.3 &  52.9 &  41.9 &  41.7 &  41.5 &  39.2 &  38.0 &  36.3 &  33.1 &  33.1 &  30.5 &  32.8 &  30.4 &  32.6 &  31.3 &  33.3 &  30.5 &  32.8 &  33.2 &  33.6 &  33.2 &     38.5 \\
Health            &  34.3 &  23.3 &  28.6 &  20.5 &  21.3 &  25.1 &  29.4 &  26.7 &  27.8 &  27.4 &  22.6 &  24.5 &  22.6 &  21.7 &  22.5 &  22.6 &  21.6 &  25.6 &  23.4 &  21.4 &  21.8 &  23.1 &  21.2 &  23.0 &     24.2 \\
Immunology        &  33.8 &  35.5 &  37.7 &  32.5 &  32.3 &  45.0 &  42.9 &  36.3 &  32.4 &  30.5 &  27.8 &  30.4 &  25.9 &  26.2 &  27.4 &  24.3 &  24.9 &  23.7 &  22.3 &  23.3 &  22.7 &  22.7 &  22.3 &  21.0 &     29.3 \\
Materials Sci.    &  56.8 &  54.7 &  52.6 &  54.8 &  50.6 &  51.6 &  47.1 &  47.2 &  44.3 &  43.5 &  42.6 &  40.5 &  39.1 &  38.5 &  35.6 &  35.5 &  34.8 &  34.5 &  34.4 &  33.5 &  33.1 &  34.2 &  35.2 &  34.5 &     42.1 \\
Mathematics       &  55.3 &  58.6 &  60.0 &  52.7 &  52.2 &  51.2 &  54.5 &  50.4 &  48.1 &  43.7 &  41.0 &  45.1 &  44.3 &  40.5 &  40.0 &  40.5 &  40.6 &  38.9 &  38.1 &  40.5 &  41.8 &  41.0 &  40.8 &  41.4 &     45.9 \\
Medicine          &  29.3 &  31.9 &  28.7 &  30.4 &  31.4 &  36.1 &  34.8 &  27.2 &  26.7 &  24.0 &  22.8 &  23.2 &  22.4 &  22.3 &  23.6 &  22.9 &  22.2 &  21.3 &  20.8 &  20.6 &  20.7 &  20.6 &  20.1 &  19.9 &     25.2 \\
Neuroscience      &  42.3 &  32.5 &  33.2 &  35.6 &  30.8 &  39.7 &  29.9 &  34.0 &  31.9 &  30.4 &  31.5 &  24.0 &  26.3 &  26.8 &  26.9 &  25.8 &  25.4 &  22.9 &  23.7 &  26.3 &  24.3 &  23.5 &  22.5 &  24.4 &     28.9 \\
Nursing           &  19.4 &  11.4 &  19.2 &  19.7 &  22.8 &  23.7 &  23.4 &  19.7 &  19.1 &  20.1 &  19.4 &  22.8 &  20.2 &  20.9 &  19.4 &  19.3 &  17.4 &  19.6 &  17.9 &  21.8 &  19.8 &  21.5 &  21.0 &  18.2 &     19.9 \\
Pharmaceutics     &  34.8 &  30.2 &  35.4 &  34.2 &  32.9 &  36.2 &  35.9 &  29.0 &  29.1 &  28.3 &  23.2 &  22.7 &  22.9 &  23.0 &  23.8 &  21.6 &  22.6 &  23.0 &  20.3 &  20.9 &  22.6 &  21.5 &  21.8 &  21.3 &     26.6 \\
Physics           &  64.0 &  61.1 &  59.2 &  58.6 &  55.8 &  55.6 &  54.4 &  50.6 &  49.5 &  50.3 &  48.7 &  47.5 &  46.5 &  43.4 &  42.5 &  42.2 &  41.0 &  41.1 &  39.6 &  38.8 &  38.8 &  39.0 &  38.2 &  36.8 &     47.6 \\
Planetary Sci.    &  58.5 &  55.6 &  55.9 &  54.2 &  56.5 &  52.5 &  50.8 &  49.7 &  49.2 &  49.5 &  50.5 &  48.8 &  49.2 &  47.3 &  47.4 &  45.2 &  42.6 &  43.4 &  42.7 &  42.4 &  41.9 &  40.7 &  41.8 &  43.1 &     48.3 \\
Psychology        &  36.9 &  28.3 &  28.5 &  29.2 &  30.1 &  43.9 &  30.2 &  36.9 &  39.7 &  28.7 &  26.9 &  28.6 &  30.3 &  28.3 &  28.2 &  28.2 &  32.6 &  29.7 &  27.5 &  29.1 &  29.4 &  29.8 &  28.2 &  29.8 &     30.8 \\
Social Sci.       &  51.0 &  40.0 &  42.4 &  39.7 &  42.0 &  40.1 &  35.6 &  45.6 &  41.3 &  40.9 &  38.5 &  35.4 &  39.3 &  34.1 &  38.0 &  38.2 &  36.6 &  36.4 &  37.8 &  38.0 &  36.8 &  37.9 &  37.9 &  38.0 &     39.2 \\
Veterinary        &  44.8 &  26.6 &  34.6 &  28.3 &  32.1 &  30.0 &  30.4 &  29.1 &  34.2 &  28.2 &  27.1 &  28.2 &  28.2 &  30.9 &  28.3 &  27.5 &  25.7 &  29.6 &  28.2 &  26.8 &  26.4 &  27.0 &  29.4 &  24.9 &     29.4 \\
\cmidrule{1-25}
All               &  29.3 &  27.8 &  28.2 &  27.0 &  26.8 &  27.5 &  28.2 &  25.1 &  24.0 &  23.8 &  23.1 &  22.4 &  22.3 &  22.4 &  23.0 &  22.4 &  22.2 &  21.9 &  21.4 &  21.1 &  20.8 &  20.7 &  20.4 &  19.8 &     23.8 \\
\bottomrule
\end{tabular}%
}
}
  \justify \small{\textit{Notes:} The table shows the share of authors with international co-affiliation on all authors with multiple affiliation by field. Shares account for countries differing shares in fields.}
\end{table}

\begin{figure}[ht!]
    \centering
    \caption{International co-affiliation by country (1996--1999 averages), network representation \label{fig:foreign_affil_network_19961999}}
    \includegraphics[width=0.6\textwidth]{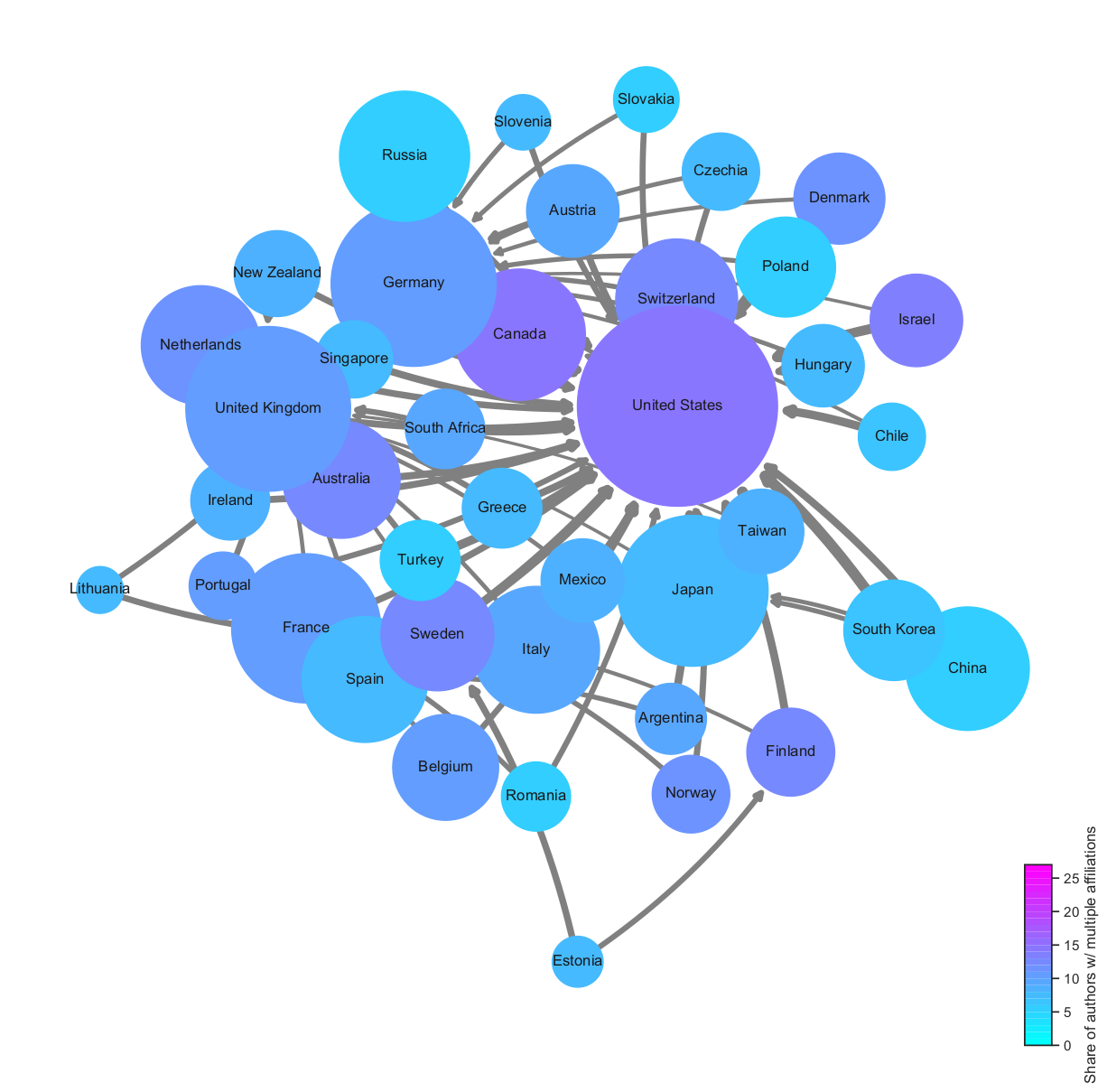}
    \justify \small{\textit{Notes:} The figure depicts a network view analogous to Figure \ref{fig:foreign_affil_shares_19961999}. Node size is proportional to the number of authors with international co-affiliation. Node color indicates the share of authors with multiple affiliations on all authors. Nodes are linked when the target country is an important host country for the international co-affiliations. We show only the two most important host countries. Edge size is proportional to the share of authors with international co-affiliation from the source country that are linked with the target country. Node position according the Fruchterman-Reingold algorithm. Graph analysis conducted with code provided by \citet{Hagberg2004NetworkX}.}
\end{figure}

\begin{figure}[ht!]
    \centering
    \caption{International co-affiliation by country (2016--2019 averages), network representation \label{fig:foreign_affil_network_20162019}}
    \includegraphics[width=0.6\textwidth]{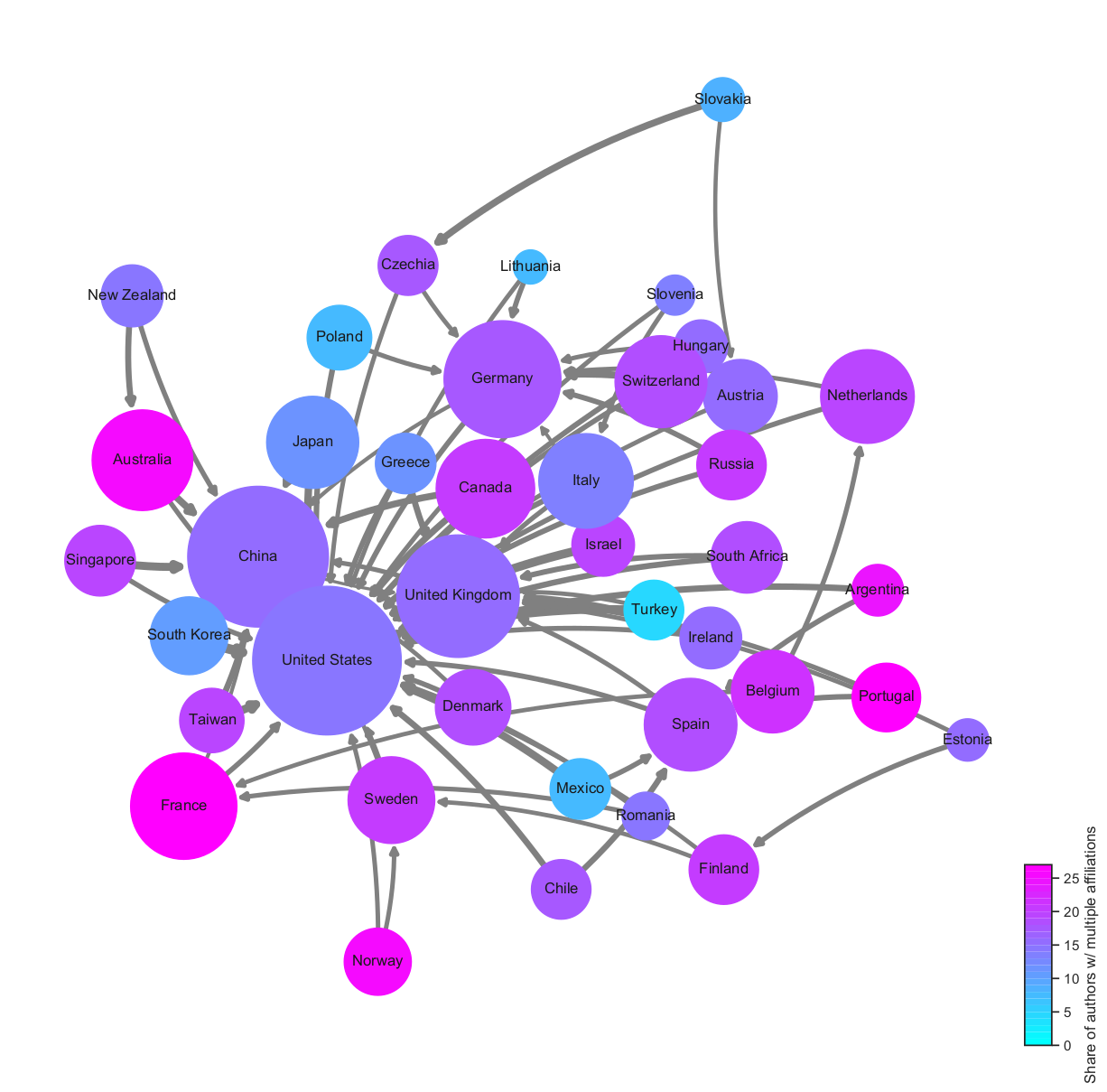}
    \justify \small{\textit{Notes:} The figure depicts a network view analogous to Figure \ref{fig:foreign_affil_shares_20162019}. Node size is proportional to the number of authors with international co-affiliation. Node color indicates the share of authors with multiple affiliations on all authors. Nodes are linked when the target country is an important host country for the international co-affiliations. We show only the two most important host countries. Edge size is proportional to the share of authors with international co-affiliation from the source country that are linked with the target country. Node position according the Fruchterman-Reingold algorithm. Graph analysis conducted with code provided by \citet{Hagberg2004NetworkX}.}
\end{figure}

\clearpage
\begin{figure}[H]
  \caption{Share of authors with multiple affiliations by country, accounting for countries' shares in fields.\label{fig:multiaff_countriesmatrix-countryfield}}
  \centering
  \includegraphics[width=\textwidth]{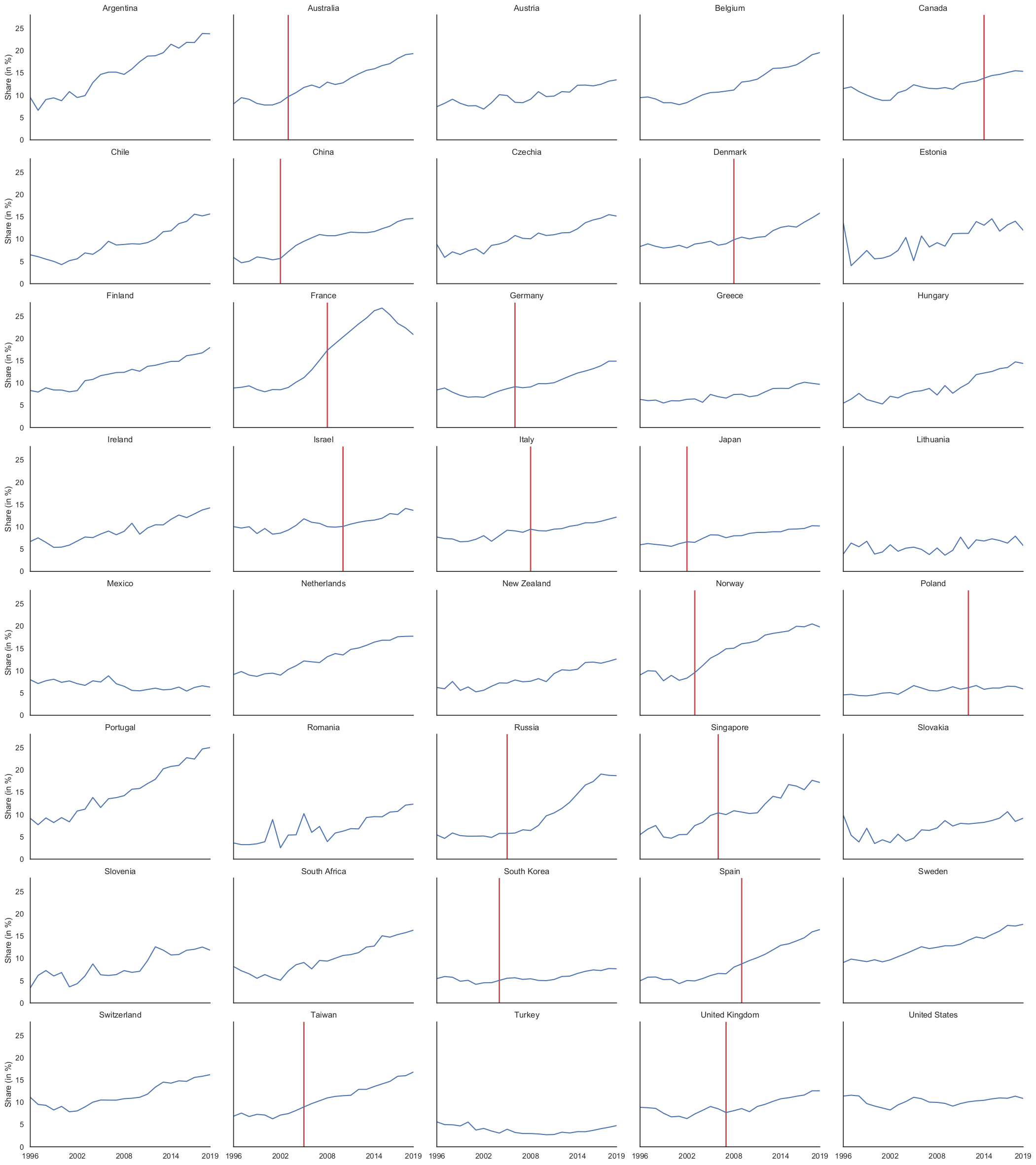}
  \justify \small{\textit{Notes:} The time series depict the share of authors with multiple affiliations $\overline{s}_{c,t}$. Red bars indicate the year of ExIns for the respective country (if applicable).}
\end{figure}

\begin{table}[ht]
 \caption{Share of authors with multiple affiliations by country accounting for countries' differing share in fields over time (in \%)
 \label{tab:multiaff_authors_share-fieldcountry}}
 \resizebox{\columnwidth}{!}{%
 \centering
\begin{tabular}{lrrrrrrrrrrrrrrrrrrrrrrrrr}
\toprule
{} &  1996 &  1997 &  1998 &  1999 &  2000 &  2001 &  2002 &  2003 &  2004 &  2005 &  2006 &  2007 &  2008 &  2009 &  2010 &  2011 &  2012 &  2013 &  2014 &  2015 &  2016 &  2017 &  2018 &  2019 &  Average \\
\midrule
Argentina      &   9.5 &   6.6 &   9.0 &   9.4 &   8.8 &  10.8 &   9.5 &   9.9 &  12.8 &  14.7 &  15.2 &  15.2 &  14.7 &  15.9 &  17.5 &  18.8 &  18.9 &  19.5 &  21.5 &  20.6 &  21.9 &  21.8 &  23.9 &  23.8 &     15.4 \\
Australia      &   8.1 &   9.5 &   9.1 &   8.2 &   7.8 &   7.8 &   8.4 &   9.7 &  10.6 &  11.8 &  12.3 &  11.7 &  12.9 &  12.4 &  12.8 &  13.9 &  14.8 &  15.6 &  15.9 &  16.7 &  17.1 &  18.3 &  19.1 &  19.3 &     12.7 \\
Austria        &   7.4 &   8.2 &   9.1 &   8.2 &   7.6 &   7.7 &   6.9 &   8.3 &  10.1 &   9.9 &   8.4 &   8.3 &   9.1 &  10.8 &   9.7 &   9.8 &  10.8 &  10.7 &  12.3 &  12.3 &  12.1 &  12.5 &  13.2 &  13.5 &      9.9 \\
Belgium        &   9.5 &   9.6 &   9.2 &   8.3 &   8.3 &   7.9 &   8.4 &   9.3 &  10.1 &  10.6 &  10.7 &  10.9 &  11.2 &  13.0 &  13.2 &  13.6 &  14.8 &  16.0 &  16.1 &  16.4 &  16.8 &  17.9 &  19.1 &  19.6 &     12.5 \\
Canada         &  11.5 &  11.9 &  10.8 &  10.0 &   9.3 &   8.9 &   8.9 &  10.6 &  11.1 &  12.4 &  11.9 &  11.5 &  11.5 &  11.7 &  11.4 &  12.6 &  12.9 &  13.2 &  13.8 &  14.4 &  14.7 &  15.1 &  15.5 &  15.4 &     12.1 \\
Chile          &   6.5 &   6.1 &   5.5 &   5.0 &   4.3 &   5.2 &   5.6 &   6.9 &   6.6 &   7.7 &   9.5 &   8.7 &   8.8 &   9.0 &   8.9 &   9.2 &  10.1 &  11.7 &  11.9 &  13.5 &  14.0 &  15.6 &  15.2 &  15.6 &      9.2 \\
China          &   5.9 &   4.7 &   5.0 &   6.0 &   5.8 &   5.3 &   5.7 &   7.2 &   8.6 &   9.5 &  10.3 &  11.0 &  10.8 &  10.8 &  11.2 &  11.6 &  11.5 &  11.4 &  11.7 &  12.3 &  12.9 &  13.9 &  14.5 &  14.6 &      9.7 \\
Czechia        &   8.8 &   5.9 &   7.1 &   6.5 &   7.4 &   7.9 &   6.7 &   8.6 &   8.9 &   9.5 &  10.8 &  10.2 &  10.1 &  11.4 &  10.8 &  11.0 &  11.4 &  11.5 &  12.3 &  13.7 &  14.3 &  14.7 &  15.5 &  15.2 &     10.4 \\
Denmark        &   8.3 &   8.9 &   8.4 &   8.0 &   8.2 &   8.6 &   8.0 &   8.9 &   9.1 &   9.5 &   8.6 &   8.9 &   9.9 &  10.4 &  10.0 &  10.4 &  10.6 &  11.9 &  12.6 &  12.9 &  12.7 &  13.8 &  14.8 &  15.8 &     10.4 \\
Estonia        &  13.6 &   4.0 &   5.7 &   7.5 &   5.6 &   5.7 &   6.3 &   7.5 &  10.4 &   5.2 &  10.7 &   8.2 &   9.2 &   8.4 &  11.2 &  11.3 &  11.3 &  13.9 &  13.1 &  14.6 &  11.8 &  13.2 &  14.0 &  12.0 &      9.8 \\
Finland        &   8.3 &   8.0 &   8.9 &   8.4 &   8.4 &   8.0 &   8.3 &  10.5 &  10.8 &  11.6 &  12.0 &  12.3 &  12.4 &  13.1 &  12.6 &  13.7 &  14.0 &  14.4 &  14.8 &  14.9 &  16.1 &  16.4 &  16.8 &  18.0 &     12.2 \\
France         &   8.9 &   9.0 &   9.4 &   8.5 &   8.0 &   8.5 &   8.5 &   9.0 &  10.2 &  11.2 &  13.0 &  15.1 &  17.3 &  18.8 &  20.3 &  21.8 &  23.3 &  24.6 &  26.2 &  26.8 &  25.3 &  23.4 &  22.4 &  20.9 &     16.3 \\
Germany        &   8.4 &   8.9 &   7.9 &   7.2 &   6.8 &   6.9 &   6.8 &   7.6 &   8.2 &   8.7 &   9.2 &   8.9 &   9.1 &   9.9 &   9.8 &  10.1 &  10.8 &  11.5 &  12.2 &  12.7 &  13.2 &  13.9 &  14.9 &  14.9 &      9.9 \\
Greece         &   6.3 &   6.0 &   6.2 &   5.5 &   6.0 &   6.0 &   6.3 &   6.4 &   5.7 &   7.4 &   6.9 &   6.6 &   7.4 &   7.5 &   6.9 &   7.2 &   8.0 &   8.8 &   8.8 &   8.8 &   9.7 &  10.2 &   9.9 &   9.7 &      7.4 \\
Hungary        &   5.5 &   6.4 &   7.7 &   6.3 &   5.8 &   5.3 &   7.0 &   6.7 &   7.5 &   8.1 &   8.3 &   8.8 &   7.3 &   9.4 &   7.7 &   9.0 &   9.9 &  11.9 &  12.2 &  12.6 &  13.2 &  13.5 &  14.7 &  14.4 &      9.1 \\
Ireland        &   6.7 &   7.5 &   6.5 &   5.4 &   5.4 &   5.9 &   6.8 &   7.7 &   7.6 &   8.4 &   9.1 &   8.2 &   9.0 &  10.8 &   8.3 &   9.7 &  10.5 &  10.4 &  11.7 &  12.7 &  12.1 &  12.9 &  13.8 &  14.3 &      9.2 \\
Israel         &  10.0 &   9.7 &  10.0 &   8.5 &   9.6 &   8.4 &   8.6 &   9.3 &  10.3 &  11.8 &  11.0 &  10.8 &  10.0 &   9.9 &  10.1 &  10.6 &  11.0 &  11.3 &  11.5 &  11.9 &  13.0 &  12.8 &  14.1 &  13.7 &     10.8 \\
Italy          &   7.7 &   7.4 &   7.3 &   6.6 &   6.7 &   7.2 &   8.0 &   6.8 &   8.0 &   9.2 &   9.1 &   8.8 &   9.5 &   9.1 &   9.1 &   9.5 &   9.6 &  10.1 &  10.4 &  10.9 &  10.9 &  11.2 &  11.7 &  12.2 &      9.0 \\
Japan          &   6.0 &   6.2 &   6.0 &   5.9 &   5.6 &   6.2 &   6.6 &   6.5 &   7.4 &   8.2 &   8.2 &   7.6 &   8.0 &   8.0 &   8.5 &   8.7 &   8.8 &   8.9 &   8.9 &   9.5 &   9.5 &   9.6 &  10.2 &  10.2 &      7.9 \\
Lithuania      &   3.9 &   6.4 &   5.5 &   6.8 &   3.9 &   4.4 &   6.0 &   4.5 &   5.2 &   5.4 &   4.9 &   3.8 &   5.3 &   3.6 &   4.7 &   7.7 &   5.1 &   7.1 &   6.8 &   7.3 &   6.9 &   6.3 &   7.9 &   5.8 &      5.6 \\
Mexico         &   8.0 &   7.1 &   7.7 &   8.1 &   7.4 &   7.7 &   7.1 &   6.7 &   7.7 &   7.5 &   8.8 &   7.1 &   6.5 &   5.6 &   5.5 &   5.8 &   6.1 &   5.7 &   5.8 &   6.3 &   5.4 &   6.2 &   6.6 &   6.3 &      6.8 \\
Netherlands    &   9.1 &   9.8 &   9.0 &   8.7 &   9.3 &   9.4 &   9.0 &  10.3 &  11.1 &  12.2 &  12.0 &  11.8 &  13.1 &  13.8 &  13.5 &  14.8 &  15.1 &  15.7 &  16.4 &  16.8 &  16.8 &  17.6 &  17.7 &  17.7 &     13.0 \\
New Zealand    &   6.2 &   6.0 &   7.6 &   5.6 &   6.3 &   5.3 &   5.6 &   6.5 &   7.3 &   7.2 &   7.9 &   7.5 &   7.6 &   8.2 &   7.5 &   9.3 &  10.2 &  10.1 &  10.3 &  11.8 &  11.9 &  11.7 &  12.1 &  12.6 &      8.4 \\
Norway         &   9.0 &  10.0 &   9.9 &   7.7 &   8.9 &   7.8 &   8.3 &   9.6 &  11.1 &  12.8 &  13.7 &  14.9 &  15.0 &  16.0 &  16.3 &  16.7 &  18.0 &  18.4 &  18.6 &  18.9 &  20.0 &  19.8 &  20.5 &  19.8 &     14.2 \\
Poland         &   4.6 &   4.7 &   4.4 &   4.4 &   4.6 &   5.0 &   5.1 &   4.7 &   5.6 &   6.7 &   6.1 &   5.6 &   5.4 &   5.8 &   6.4 &   5.9 &   6.2 &   6.7 &   5.8 &   6.1 &   6.1 &   6.5 &   6.5 &   5.9 &      5.6 \\
Portugal       &   9.1 &   7.7 &   9.2 &   8.2 &   9.3 &   8.4 &  10.8 &  11.2 &  13.8 &  11.6 &  13.5 &  13.8 &  14.2 &  15.7 &  15.8 &  16.9 &  17.9 &  20.2 &  20.8 &  21.0 &  22.8 &  22.4 &  24.7 &  25.0 &     15.2 \\
Romania        &   3.6 &   3.3 &   3.2 &   3.4 &   3.9 &   8.9 &   2.5 &   5.4 &   5.4 &  10.2 &   6.0 &   7.3 &   3.9 &   5.8 &   6.3 &   6.8 &   6.8 &   9.3 &   9.5 &   9.5 &  10.5 &  10.7 &  12.1 &  12.3 &      7.0 \\
Russia         &   5.4 &   4.6 &   5.9 &   5.3 &   5.1 &   5.2 &   5.2 &   4.9 &   5.8 &   5.8 &   5.9 &   6.6 &   6.4 &   7.6 &   9.7 &  10.4 &  11.4 &  12.8 &  14.7 &  16.6 &  17.4 &  19.1 &  18.8 &  18.7 &      9.5 \\
Singapore      &   5.5 &   6.8 &   7.5 &   5.0 &   4.7 &   5.5 &   5.5 &   7.5 &   8.2 &   9.8 &  10.4 &  10.0 &  10.9 &  10.5 &  10.2 &  10.4 &  12.4 &  14.1 &  13.7 &  16.7 &  16.4 &  15.6 &  17.7 &  17.2 &     10.5 \\
Slovakia       &   9.9 &   5.4 &   3.8 &   6.9 &   3.5 &   4.3 &   3.7 &   5.6 &   4.0 &   4.7 &   6.6 &   6.4 &   7.0 &   8.6 &   7.4 &   8.0 &   7.9 &   8.1 &   8.2 &   8.6 &   9.2 &  10.6 &   8.5 &   9.2 &      6.9 \\
Slovenia       &   3.4 &   6.2 &   7.3 &   6.1 &   6.8 &   3.6 &   4.3 &   6.1 &   8.8 &   6.3 &   6.2 &   6.4 &   7.3 &   6.9 &   7.1 &   9.5 &  12.6 &  11.9 &  10.8 &  10.9 &  11.8 &  12.0 &  12.6 &  11.8 &      8.2 \\
South Africa   &   8.2 &   7.2 &   6.5 &   5.5 &   6.4 &   5.6 &   5.1 &   7.2 &   8.6 &   9.1 &   7.6 &   9.5 &   9.4 &  10.0 &  10.7 &  10.9 &  11.3 &  12.5 &  12.8 &  15.1 &  14.8 &  15.3 &  15.8 &  16.3 &     10.1 \\
South Korea    &   5.5 &   5.9 &   5.8 &   4.9 &   5.1 &   4.2 &   4.5 &   4.6 &   5.1 &   5.5 &   5.7 &   5.3 &   5.4 &   5.1 &   5.0 &   5.3 &   5.9 &   6.0 &   6.6 &   7.1 &   7.4 &   7.3 &   7.7 &   7.7 &      5.8 \\
Spain          &   5.0 &   5.8 &   5.8 &   5.2 &   5.3 &   4.3 &   5.0 &   5.0 &   5.5 &   6.1 &   6.6 &   6.5 &   8.0 &   8.8 &   9.5 &  10.2 &  10.9 &  11.9 &  12.9 &  13.3 &  13.9 &  14.6 &  15.9 &  16.5 &      8.9 \\
Sweden         &   9.1 &   9.8 &   9.6 &   9.3 &   9.7 &   9.2 &   9.7 &  10.4 &  11.1 &  11.8 &  12.6 &  12.2 &  12.5 &  12.8 &  12.8 &  13.2 &  14.1 &  14.8 &  14.5 &  15.4 &  16.1 &  17.4 &  17.3 &  17.6 &     12.6 \\
Switzerland    &  11.1 &   9.5 &   9.3 &   8.3 &   9.1 &   7.9 &   8.1 &   9.0 &  10.0 &  10.5 &  10.5 &  10.5 &  10.8 &  10.9 &  11.1 &  11.8 &  13.4 &  14.5 &  14.3 &  14.8 &  14.7 &  15.6 &  15.8 &  16.2 &     11.6 \\
Taiwan         &   6.9 &   7.6 &   6.8 &   7.3 &   7.1 &   6.3 &   7.1 &   7.4 &   8.2 &   8.9 &   9.7 &  10.3 &  11.0 &  11.3 &  11.5 &  11.6 &  12.9 &  12.9 &  13.5 &  14.1 &  14.7 &  15.8 &  15.9 &  16.8 &     10.7 \\
Turkey         &   5.6 &   5.0 &   4.9 &   4.7 &   5.6 &   3.8 &   4.1 &   3.5 &   3.1 &   3.9 &   3.2 &   3.0 &   3.0 &   2.9 &   2.7 &   2.8 &   3.3 &   3.1 &   3.4 &   3.4 &   3.7 &   4.1 &   4.4 &   4.8 &      3.8 \\
United Kingdom &   8.9 &   8.8 &   8.6 &   7.5 &   6.7 &   6.8 &   6.3 &   7.4 &   8.2 &   9.0 &   8.5 &   7.7 &   8.1 &   8.6 &   7.9 &   9.1 &   9.5 &  10.2 &  10.8 &  11.0 &  11.3 &  11.6 &  12.6 &  12.6 &      9.1 \\
United States  &  11.4 &  11.6 &  11.4 &   9.7 &   9.1 &   8.7 &   8.3 &   9.4 &  10.2 &  11.1 &  10.8 &  10.0 &  10.0 &   9.7 &   9.1 &   9.7 &  10.1 &  10.3 &  10.4 &  10.8 &  11.0 &  10.9 &  11.4 &  10.9 &     10.2 \\
\bottomrule
\end{tabular}%
}
\justify \small{\textit{Notes:} The table shows the share of authors $\overline{s}_{c,t}$ with multiple affiliations by country and year accounting for countries' differing shares in fields. In each year we count unique authors based on their Scopus Author ID.}
\end{table}

\begin{table}[ht]
 \caption{Share of authors with multiple affiliations by country group accounting for countries' differing share in fields over time (in \%)
 \label{tab:multiaff_groups-countryfield}}
 \resizebox{\columnwidth}{!}{%
 \centering
\begin{tabular}{lrrrrrrrrrrrrrrrrrrrrrrrr}
\toprule
{} &  1996 &  1997 &  1998 &  1999 &  2000 &  2001 &  2002 &  2003 &  2004 &  2005 &  2006 &  2007 &  2008 &  2009 &  2010 &  2011 &  2012 &  2013 &  2014 &  2015 &  2016 &  2017 &  2018 &  2019 \\
\midrule
Control group  &   7.9 &   7.1 &   7.4 &   7.0 &   6.9 &   6.9 &   6.8 &   7.7 &   8.6 &   8.9 &   9.2 &   9.0 &   9.1 &   9.7 &   9.6 &  10.5 &  11.0 &  11.8 &  12.1 &  12.7 &  12.9 &  13.5 &  14.0 &  14.0 \\
ExIn countries &   7.2 &   6.8 &   7.0 &   6.8 &   6.4 &   6.5 &   6.5 &   7.2 &   8.2 &   8.8 &   9.6 &  10.4 &  10.9 &  11.8 &  12.8 &  13.4 &  14.2 &  15.1 &  16.2 &  17.1 &  17.2 &  17.6 &  17.6 &  17.3 \\
\bottomrule
\end{tabular}%
}
\justify \small{\textit{Notes:} The table shows the share of authors $\overline{s}_{c,t}$ with multiple affiliations by country and year accounting for countries' differing shares in fields by country group.}
\end{table}

\begin{figure}[H]
  \caption{Share of authors with multiple affiliation by field over time accounting for countries' differing share in fields, excluding authors based in China\label{fig:multiaff_fields-countryfield_nochina}}
  \centering
  \includegraphics[width=0.57\textwidth]{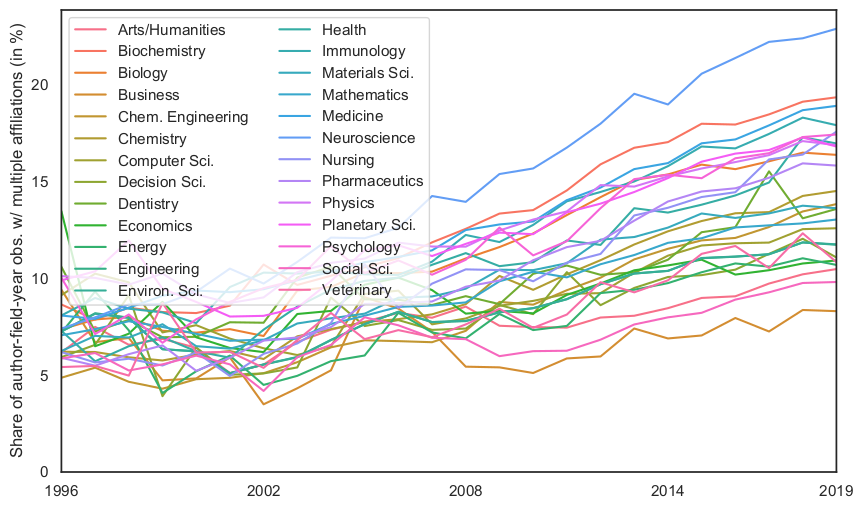}
  \justify \small{\textit{Notes:} Figure is analogous to Figure \ref{fig:multiaff_fields-countryfield} but excludes authors whose first affiliation is based in China.}
\end{figure}

\FloatBarrier

\section{Data cleaning and aggregation\label{sec::data}}

We use data from the extensive Scopus database with two corrections. First, we do not treat joint affiliation to a university system and to a member of that system as multiple affiliation. For example, someone affiliated both with the "University of California" and "University of California, Los Angeles" does not count as author having multiple affiliations.

Second, we also remove co-affiliations with the IEEE and the IEEE Canada. These are affiliations where authors state their membership when publishing in corresponding journals, but Scopus often confuses them with an affiliation to an organisation. There might be further measurement error due to new affiliations appearing in the data, which Scopus does not recognise as an institution and therefore cannot assign to an organisation type. Accuracy in recognizing affiliations may therefore decline as a consequence of new affiliations appearing.

\begin{figure}[ht]
  \centering
  \caption{Affiliations most commonly listed in multiple affiliations over time\label{fig:multiaffs_top}}
  \includegraphics[width=0.6\textwidth]{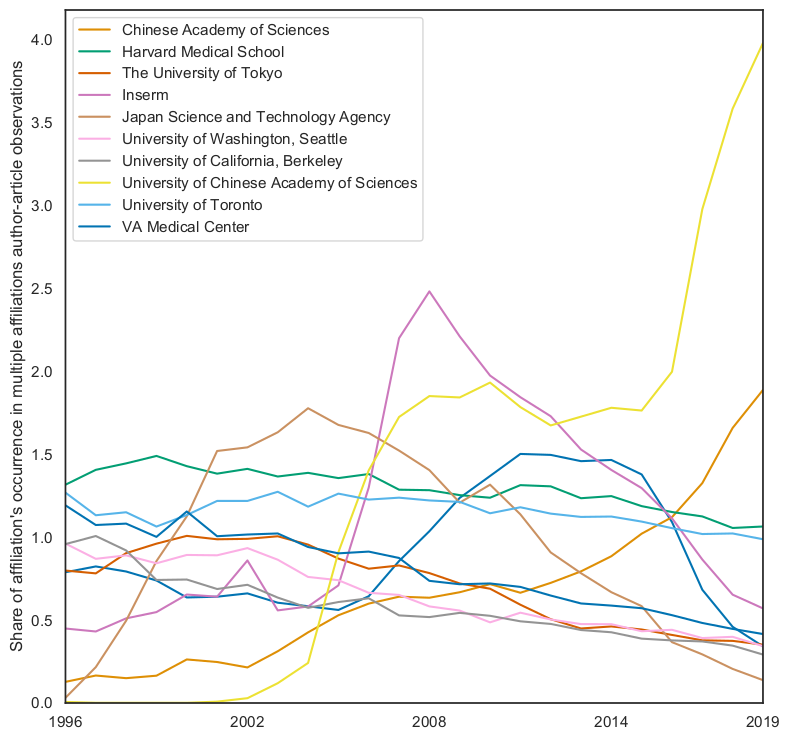}
  \justify \small{\textit{Notes:} How often an affiliation is part of a multiple affiliation combination, as share of all multiple affiliation author-article observations. The figure only considers affiliations that are among the top four mentioned affiliation in at least one year.}
\end{figure}

Figure \ref{fig:multiaffs_top} plots how often an affiliation is part of a multiple affiliation, as share of all author-article observations with multiple affiliations. The figure shows that that there is indeed some concentration of affiliations occurring with specific institutions and that their importance increased over time. The most prominent institutions are in China with the Chinese Academy of Sciences and the Universities of the Chinese Academy of Sciences. In France, Inserm (the institute national de la santé et de la recherche médicale is the French National Institute of Health and Medical Research) stands out. Overall, even these very frequently named affiliations constitute never more than 5\% of co-affiliations globally.

\end{document}

%% file: header.tex
\usepackage[utf8]{inputenc} \usepackage[T1]{fontenc}
\usepackage{times}

\usepackage{natbib}

\usepackage[usenames, dvipsnames, svgnames, table]{xcolor}
\usepackage[plainpages=false]{hyperref}
\hypersetup{colorlinks=true, urlcolor=Cerulean, citecolor=Cerulean, linkcolor=Cerulean}

\usepackage{authblk}
\usepackage{pdfpages, placeins}
\usepackage{ragged2e}
\usepackage{enumitem}
\usepackage{pdflscape}
\usepackage{setspace}
\AtBeginDocument{\lefthyphenmin=3 }

\usepackage{geometry}
\usepackage{booktabs,array,dcolumn}
\newcommand*\ctoprule[1]{\addlinespace\cmidrule[\heavyrulewidth]{#1}}
\newcommand*\cbottomrule[1]{\cmidrule[\heavyrulewidth]{#1}\addlinespace}
\newcolumntype{d}[1]{D{.}{.}{#1} }

\usepackage[caption = false]{subfig}
\usepackage{graphicx,afterpage}

\usepackage{float}
\usepackage{amsmath}
\usepackage{amssymb}

\usepackage{endnotes}
\let\footnote=\endnote

%% file: multiaff_statistics.tex
\newcommand{\NOfArticlesUnique}{22,198,910}
\newcommand{\NOfAuthorcountryfieldyear}{111,038,114}
\newcommand{\NOfAuthorcountryyear}{48,352,112}
\newcommand{\NOfAuthorfieldyear}{109,448,897}
\newcommand{\NOfAuthoroctileyear}{63,935,940}
\newcommand{\NOfAuthorpaper}{41,119,839}
\newcommand{\NOfAuthorsUnique}{15,234,353}